\documentclass{JHEP3}
\usepackage{epsfig, multicol, bbm, amsmath, amssymb, euscript, array, mathrsfs, amsfonts}

\title{Unified Dark Matter models with fast transition}

\author{Oliver F. Piattella$^{a,b,c}$, Daniele Bertacca$^{c,d,e}$, Marco Bruni$^{c}$ and Davide Pietrobon$^{c,f}$\\
$^a$ Dipartimento di Scienze Fisiche e Matematiche, Universit\`a dell'Insubria, Via Valleggio 11, 22100 Como, Italy\\
$^b$ INFN, sez. di Milano, Via Celoria 16, 20133 Milano, Italy\\
$^c$ Institute of Cosmology and Gravitation, University of Portsmouth,
Dennis Sciama Building, Portsmouth PO1 3FX United Kingdom\\
$^d$ Dipartimento di Fisica Galileo Galilei  Universit\`{a} di Padova\\
$^e$ INFN Sezione di Padova, via F. Marzolo, 8 I-35131 Padova, Italy\\
$^f$ Dipartimento di Fisica, Universit\`{a} di Roma “Tor Vergata”, via della Ricerca Scientifica 1, 00133 Roma, Italy\\
E-mails: \email{oliver.piattella@uninsubria.it}, \email{bertacca@pd.infn.it}, \email{marco.bruni@port.ac.uk}, \email{davide.pietrobon@roma2.infn.it}}

\preprint{\arXivid{0911.2664}}

\abstract{We investigate the general properties of Unified Dark Matter (UDM)  fluid models where the pressure and the energy density are linked by a barotropic equation of state (EoS)  $p=p(\rho)$ and the perturbations are adiabatic. The EoS is assumed to admit a future attractor that acts as an effective cosmological constant, while asymptotically in the past the pressure is negligible. UDM models of the dark sector are appealing because they evade the so-called ``coincidence problem'' and ``predict" what can be interpreted as $w_{\rm DE} \approx -1$, but  in general suffer the effects of  a non-negligible Jeans scale that wreak havoc  in the evolution of perturbations, causing a large Integrated Sachs-Wolfe effect and/or changing structure formation at small scales.  Typically,   observational constraints are violated, unless the  parameters of the UDM model are tuned to make it indistinguishable from $\Lambda$CDM. Here we show how this  problem can be avoided, studying in detail  the functional form  of the Jeans scale in adiabatic UDM perturbations and  introducing a class of models with a fast transition between an early Einstein--de\! Sitter CDM-like era and a later $\Lambda$CDM-like phase. If the transition is fast enough, these models  may exhibit  satisfactory  structure formation and CMB fluctuations. To consider a concrete case, we introduce a toy UDM model and show that it can predict CMB and matter power spectra that are in agreement with observations for a wide range of parameter values.}


\begin{document}

\section{Introduction}

In the last three decades the flat $\Lambda$CDM model \cite{Peebles:1984ge,Efstathiou:1990xe} has emerged as the  standard ``concordance'' \cite{Spergel:2003cb,Tegmark:2003ud} model of cosmology. It assumes General Relativity (GR) as the correct theory of gravity, with two unknown components  dominating the dynamics of the  late Universe: {\it i )} a cold collisionless Cold Dark Matter (CDM) describing some weakly interacting particles, responsible for structure
formation, {\it ii )} a cosmological constant $\Lambda$\ \cite{Weinberg:1989, Sahni:1999gb} making up the balance to make the Universe spatially flat and  driving the observed cosmic acceleration \cite{Riess:1998cb,Perlmutter:1998np,Riess:1998dv,Kowalski:2008ez}. The main alternative to the cosmological constant is a more general dynamic component called Dark Energy (DE) \cite{Copeland:2006wr,Peebles:2002gy,Padmanabhan:2002ji}. 
Many independent observations support both the existence of a CDM component and that of a separate DE \cite{Kowalski:2008ez,Allen:2004cd,Tegmark:2006az,Percival:2006kh,Percival:2007yw,Komatsu:2008hk,Dunkley:2008ie,Hinshaw:2008kr,Percival:2009xn,Reid:2009xm}.

Early proposals \cite{Peebles:1984ge,Efstathiou:1990xe} of the $\Lambda$CDM model were adding $\Lambda$ to CDM in an attempt to conciliate in the simplest possible way the emerging inflationary paradigm, which requires a spatially flat Universe and an almost scale-invariant spectrum of perturbations, with the observed low density of matter.  
It should however be recognised that, while some form of CDM is independently expected to exist within any modification of the Standard Model of high energy physics, the really compelling reason to postulate the existence of DE has been the  cosmic acceleration measured in the last decade \cite{Riess:1998cb,Perlmutter:1998np,Riess:1998dv,Kowalski:2008ez,Percival:2007yw,Komatsu:2008hk,Dunkley:2008ie,Hinshaw:2008kr,Percival:2009xn}.
It is mainly for this reason that it is worth investigating the hypothesis that CDM and DE are the two faces of a single Unified Dark Matter (UDM) component, thereby also   avoiding the so-called ``coincidence problem'' \cite{Zlatev:1998tr}.

Other attempts to explain the observed acceleration also exist, most notably by assuming a gravity theory other than GR, or an interaction between DM and DE (see e.g.\ \cite{Copeland:2006wr, Durrer:2008, Koyama:2007rx, Capozziello:2007ec, Sotiriou:2008rp} and \cite{Quercellini:2008vh, Valiviita:2008iv, CalderaCabral:2008bx, Majerotto:2009np, CalderaCabral:2009ja, Valiviita:2009nu, Bassett:2002fe}). In this paper however we focus on UDM models, where this single matter component  provides an  explanation for structure formation and cosmic acceleration. 

In general, in the $\Lambda$CDM model or in most  models with DM and DE, the CDM component is free to form structures at all scales, with  DE  only affecting the general overall expansion \cite{Copeland:2006wr,Peebles:2002gy,Padmanabhan:2002ji}.
Instead, a general feature of UDM models  is the possible appearance of an effective sound speed, which may become  significantly different from zero during the Universe evolution, then corresponding in general to the appearance of a Jeans length (i.e.\ a sound horizon) below which the dark fluid does not cluster (e.g.\ see \cite{Hu:1998kj, Bertacca:2007cv, Pietrobon:2008js}). Moreover, the presence of a non-negligible speed of sound can modify the evolution of the gravitational potential, producing a strong Integrated Sachs Wolfe (ISW) effect \cite{Bertacca:2007cv}. Therefore, in UDM models it is crucial to study the evolution of the effective speed of sound and that of the Jeans length.

Several adiabatic fluid models and models based on non canonical kinetic Lagrangians have been investigated in the literature. For example, the generalised Chaplygin gas \cite{Kamenshchik:2001cp, Bilic:2001cg, Bento:2002ps} (see also \cite{Makler:2002jv, Bento:2002yx, Alcaniz:2002yt, Santos:2006ce, Gorini:2007ta, Sen:2005sk, Makler:2003iw, Carturan:2002si, Amendola:2003bz, Sandvik:2002jz, Giannantonio:2006ij}), the Scherrer \cite{Scherrer:2004au} and generalised Scherrer solutions \cite{Bertacca:2007ux}, the single dark perfect fluid with ``affine'' 2-parameter barotropic equation of state (see \cite{Balbi:2007mz,Pietrobon:2008js} and the corresponding scalar field models \cite{Quercellini:2007ht}) and the homogeneous scalar field deduced from the galactic halo space-time \cite{DiezTejedor:2006qh,Bertacca:2007fc}.
In general, in order for UDM models to  have a background evolution that fits observations and a very small speed of sound, a severe fine-tuning of their parameters is necessary (see for example \cite{Pietrobon:2008js, Makler:2003iw, Carturan:2002si, Amendola:2003bz, Sandvik:2002jz, Scherrer:2004au, Giannakis:2005kr, Piattella:2009da}).
Finally, one could also easily reinterpret UDM models based on a scalar field Lagrangian in terms of generally non-adiabatic fluids \cite{DiezTejedor:2005fz, Brown:1992kc} (see also \cite{Bertacca:2007ux, Bertacca:2008uf}). For these models the effective speed of sound, which remains defined in the context of linear perturbation theory, is not the same as the adiabatic speed of sound (see \cite{Hu:1998kj}, \cite{Garriga:1999vw} and \cite{Mukhanov:2005sc}).
In \cite{Bertacca:2008uf} a reconstruction technique is devised for the Lagrangian, which allows  to find models where the effective speed of sound is small enough, such that the {\it k}-essence scalar field can cluster (see also \cite{Camera:2009uz}).

In the present paper we investigate the possibility of constructing adiabatic UDM models where the Jeans length is very small, even when the speed of sound is not negligible. In particular, our study is focused on models that admit an effective cosmological constant and that are characterised by a short period during which the effective speed of sound varies significantly from zero. This  allows a fast transition between an early matter dominated era, which is indistinguishable from an Einstein--de\! Sitter model, and a more recent epoch whose dynamics, background and perturbative,  are very close to that of a standard $\Lambda$CDM model.

To consider a concrete example, we introduce a 3-parameter class of  toy UDM adiabatic models with fast transition.  One of the parameters is the effective cosmological constant $\rho_{\Lambda}$ or, equivalently, the corresponding density parameter $\Omega_{\Lambda}$; the other two are $\rho_{\rm s}$ and $\rho_{\rm t}$, respectively regulating how fast the transition is and the redshift of the transition. Studying the Jeans scale in these models we find  an approximate analytical relation that sets a constraint on these two  parameters, a sufficient condition that  $\rho_{\rm s}$ and $\rho_{\rm t}$ have to satisfy in order for the models to be minimally viable.  This relation can be used to fix $\rho_{\rm s}$ for any given $\rho_{\rm t}$: in this case, with respect to a flat $\Lambda$CDM, in practice our models have one single extra parameter. With the help of this relation we establish our  main result:  if the fast transition takes place early enough, at a redshift $z\gtrsim 2$ when the effective cosmological constant is still subdominant, then the predicted background evolution,  Cosmic Microwave Background (CMB) anisotropy and  linear matter power spectrum are in agreement  with observations for a broad range of parameter values.  In practice, in our toy models the predicted CMB  and matter power spectra  do not display significant differences from those computed in the $\Lambda$CDM model, because the Jeans length remains  small at all times, except for negligibly short periods,  even if during the fast transition the speed of sound can be large. In other words,  this kind of adiabatic UDM models  evade the  ``no-go theorem'' of Sandvik {\it et al} \cite{Sandvik:2002jz} who, studying the generalised Chaplygin gas UDM models,  showed that this broad class must have an almost constant negative pressure at all times  in order to satisfy observational constraints,  making these models  in practice indistinguishable from the $\Lambda$CDM model (see also \cite{Pietrobon:2008js}).

The paper is organised as follows: in section \ref{sec:bgperteq} we introduce the basic equations describing the background and the perturbative evolution. In section \ref{sec:eos} we use the pressure-density plane to analyse  the properties that a general barotropic UDM model has to fulfil in order to be viable. In section \ref{sec:tghmodel} we introduce our  toy UDM model with fast transition and study its background evolution, comparing it  to a $\Lambda$CDM. In section \ref{sec:perts} we analyse the properties of perturbations in this model, focusing on the the evolution of the  effective speed of sound  and that of the Jeans length during the  transition. 
Then, using the CAMB code \cite{Lewis:1999bs}, in section \ref{sec:spectra} we compute the CMB and the matter power spectra. Finally, section \ref{sec:concl} is devoted to our conclusions.

\section{Background and perturbative equations for a UDM model}\label{sec:bgperteq}

We assume a spatially flat Friedmann-Lema\^{\i}tre-Robertson-Walker (FLRW) cosmology. The metric
 then is $ds^2 = -dt^2 + a^2(t)\delta_{ij}dx^idx^j$, where $t$ is the cosmic time, $a(t)$ is the scale factor and $\delta_{ij}$ is the Kronecker delta. The total stress-energy tensor is that of a  perfect fluid: $T_{\mu\nu} = \left(\rho + p\right)u_{\mu}u_{\nu} + pg_{\mu\nu}$, where $\rho$ and $p$ are, respectively, the energy density and the pressure of the fluid, while $u^{\mu}$ is its four-velocity.
Starting from these assumptions, and choosing units such that $8\pi G = c = 1$, Einstein equations imply the Friedmann and Raychaudhuri equations:
\begin{eqnarray}\label{Fried}
 H^2 & =& \left(\frac{\dot{a}}{a}\right)^2 = \frac{\rho}{3}\;,\\
\label{Raycha}
 \frac{\ddot{a}}{a} &=& -\frac{1}{6}\left(\rho + 3p\right)\;,
\end{eqnarray}
where $H=\dot{a}/a$ is the Hubble expansion scalar and the dot denotes derivative with respect to the cosmic time. Assuming that the energy density of the radiation is negligible at the times of interest, and disregarding also the small baryonic component, $\rho$ and $p$ represent the energy density and the  pressure of the UDM component.

The energy conservation equation is:
\begin{equation}\label{enconseq}
 \dot{\rho} = -3H\left(\rho + p\right) = -3H\rho\left(1 + w\right)\;,
\end{equation}
where $w := p/\rho$ is the equation of state (hereafter EoS) ``parameter''. In this paper we investigate a class of UDM models based on a barotropic EoS $p = p(\rho)$, i.e.\ those models for which the pressure is function of the density only (see e.g.\ \cite {Ananda:2005xp} and \cite{Linder:2008ya} for a discussion of general properties of barotropic fluids as dark components). In this case, if the EoS allows the value $w = -1$, the barotropic fluid admits  an effective cosmological constant energy density, i.e.\ a fixed point of Eq.\ (\ref{enconseq}) \cite{Ananda:2005xp,Ananda:2006gf,Balbi:2007mz},  which we will denote as $\rho_\Lambda$. Under very reasonable conditions (see the discussion below) this effective cosmological constant is unavoidable for barotropic fluids\footnote{Since this effective cosmological constant trivially satisfies an energy conservation equation (\ref{enconseq}) on its own, a fluid admitting an effective $\rho_{\Lambda}$ is  always equivalent to two separate components, namely $\rho_{\Lambda}$ itself and an ``aether'' fluid, see \cite {Ananda:2005xp} and \cite{Linder:2008ya}. Obviously, this is more general; for instance, a scalar field with a potential admitting a non vanishing minimum $V_{0}=V(\phi_{0})\not =0$ at - say - $\phi_{0}$, is equivalent to a cosmological constant $\rho_{\Lambda}=V_{0}$ and a scalar field in a potential $\tilde{V}=V-V_{0}$.} \cite{Ananda:2005xp,Ananda:2006gf}.

In order to properly describe the dynamics of the fluid we must consider the background EoS as well as the speed of sound which regulates the growth of fluid perturbations on different cosmological scales. In the following we shall confine our study to the simplest hypothesis that the EoS remains of the barotropic form $p=p(\rho)$ when we allow for perturbations: in this case our models will be adiabatic, and the effective and adiabatic speeds of sound will coincide, see e.g.\  \cite{Bardeen:1980kt,Kodama:1985bj,Bruni:1992dg,Hu:1998kj}. Other choices for the perturbed spacetime are possible, see \cite{Pietrobon:2008js} for a recent practical example.

Let us consider small perturbations of the FLRW metric in the longitudinal gauge, using conformal time $\eta$: $ds^2 = -a^2(\eta)\left[\left(1 + 2\Phi\right)d\eta^2 - \left(1 - 2\Phi\right)\delta_{ij}dx^idx^j\right]$, where $\Phi$ is the gravitational potential.

Defining
\begin{equation}
u := \frac{2\Phi}{\sqrt{\rho + p}}\;
\end{equation}
and linearising the 0-0 and 0-i components of Einstein equations, for a plane-wave perturbation $u \propto \exp\left(i{\bf k} \cdot {\bf x}\right)$ one obtains the following second order differential
equation \cite{Mukhanov:1990me,Giannakis:2005kr,Bertacca:2007cv}:
\begin{equation}\label{equ}
u'' + k^{2}c_{\rm s}^{2}u - \frac{\theta''}{\theta}u = 0\;,
\end{equation}
where the prime is the derivative with respect to the conformal time $\eta$, $c_{\rm s}^2$ is the effective speed of sound and 
\begin{equation}\label{theta}
 \theta := \sqrt{\frac{\rho}{3(\rho + p)}}(1 + z)\;,
\end{equation}
with  $z$  the redshift, $1 + z = a^{-1}$. In general, the adiabatic speed of sound is $c_{\rm ad}^2 := p'/\rho'$; for an adiabatic fluid  $c_{\rm s}^2 = c_{\rm ad}^2$.

Starting from Eq. (\ref{equ}), let us define the squared Jeans wave number \cite{Bertacca:2007cv}:
\begin{equation}
 k^{2}_{\rm J} := \left|\frac{\theta''}{c_{\rm s}^{2}\theta}\right|\;.
\end{equation}
Its reciprocal  defines the squared Jeans length: $\lambda^2_{\rm J} := a^{2}/k^2_{\rm J}$.

There are two regimes of evolution. If $k^2 \gg k_{\rm J}^2$ and the speed of sound is slowly varying, then the solution of Eq. (\ref{equ}) is
\begin{equation}
 u \simeq \frac{C}{\sqrt{c_{\rm s}}}\exp\left(\pm ik\int c_{\rm s}d\eta\right)\;,
\end{equation}
where $C$ is an appropriate integration constant\footnote{This solution is exact if the speed of sound satisfies the equation $2c_{\rm s}''c_{\rm s} - 3\left(c_{\rm s}'\right)^2 = 0$, which implies 
\begin{eqnarray}
 c_{\rm s} = \frac{4}{\left(c_1\eta + c_2\right)^2}\;,\nonumber
\end{eqnarray}
where $c_1$ and $c_2$ are generic constants. A particular case is when $c_1 = 0$, for which the speed of sound is constant.}.
On these scales, smaller than the Jeans length, the gravitational potential oscillates and decays in  time, with observable effects on both the CMB and the matter power spectra \cite{Bertacca:2007cv}.

For large scale perturbations, when $k^2 \ll k_{\rm J}^2$, Eq. (\ref{equ}) can be rewritten as  $u''/u  \simeq \theta''/\theta$, with  general solution
\begin{equation}\label{uklesskJ}
 u \simeq \kappa_1\theta + \kappa_2\theta\int \frac{d\eta}{\theta^{2}}\;.
\end{equation}
In this large scale limit the evolution of the gravitational potential $\Phi$ depends only on the background evolution, encoded in $\theta$, i.e.\ it is the same for all $k$ modes. The first term $\kappa_1\theta$ is the usual decaying mode, which we are going to neglect in the following, while $\kappa_2$ is related to the power spectrum, see e.g.\ \cite{Mukhanov:2005sc}. 

\section{Analysis of barotropic UDM models on the pressure-density plane}\label{sec:eos}
\begin{figure}[htbp]
\begin{center}
\includegraphics[width=0.6\columnwidth]{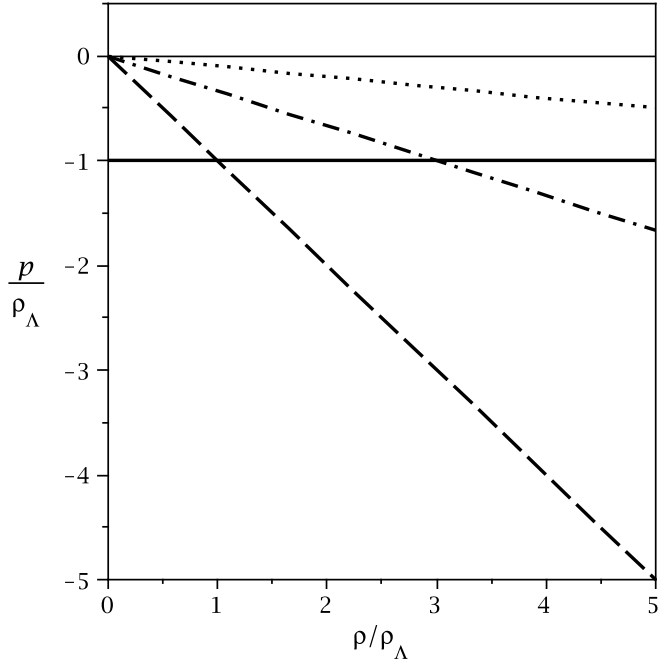}
\caption{The UDM  $p-\rho$ plane with the most important areas, (see the text for more detail). The dashed line represents the $p = -\rho$ line; the dash-dotted line represents the $p = -\rho/3$ line, the boundary between the decelerated expansion phase of the Universe and the accelerated one; the dotted line  $p = -\rho/10$ represents a fictitious boundary, above which  the CDM-like behaviour of the UDM fluid dominates. The pressure and the energy density are normalised to $\rho_\Lambda$. The $\Lambda$CDM model is represented here by the solid horizontal line $p/\rho_\Lambda=-1$, while the line $p=0$ represents an EdS model, i.e. pure CDM.}
\label{fig:p_rho}
\end{center}
\end{figure}
A common way  to study the properties of the EoS of DE  is to consider the $(dw/d\ln a) - w$ phase space (see e.g.\ \cite{Caldwell:2005tm,Scherrer:2005je,dePutter:2007ny,Chen:2009bca}).
Here we follow another approach, studying our models in the pressure-density plane, see  Fig.\ \ref{fig:p_rho}. There are several motivations for this choice. First of all, in the barotropic case  we are  considering  the pressure is a function of the density only, so it is natural to give a graphical description on the $p-\rho$ plane. Second, this plane gives an idea of the cosmological evolution of the dark fluid. Indeed, in an expanding Universe ($H>0$) Eq.\ 
 (\ref{enconseq}) implies $\dot{\rho}<0$ for a fluid satisfying the null energy condition \cite{Visser:1997aa} $w> -1$ during its evolution, hence there exists a one-to-one correspondence between time and energy density. 
Finally, in the adiabatic case  the effective speed of sound we  have introduced in Eq. (\ref{equ}) can be  written as $c_{\rm s}^2 = dp/d\rho$, therefore it has an immediate geometric significance on the $p-\rho$ plane as the slope of the curve describing the EoS $p=p(\rho)$. 

For a fluid, it is quite natural to assume   $c_{\rm s}^2\geq 0$, which then  implies that the function $p(\rho)$ is monotonic, and as such crosses the $p = -\rho$ line at some point $\rho_{\Lambda}$.\footnote{Obviously, we are assuming that during the evolution the EoS allows $p$ to  become negative, actually violating  \ the strong energy condition \cite{Visser:1997aa}, i.e.\  $p< -\rho/3$ at least for some $\rho>0$, otherwise the fluid would never be able to produce an accelerated expansion.} From the point of view of the dynamics  this is a crucial fact, because it implies  the existence of an attracting  fixed point ($\dot{\rho}=0$) for the conservation equation (\ref{enconseq}) of our UDM fluid, i.e.\  $\rho_{\Lambda}$ plays the role of an unavoidable effective cosmological constant. The Universe necessarily evolves toward  an asymptotic de-Sitter phase, a sort of cosmic no-hair theorem  (see  \cite{Bruni:2001pc, Bruni:1994cv} and refs.\ therein and \cite{Ananda:2005xp,Ananda:2006gf,Balbi:2007mz}). 

We now summarise, starting from Eqs.\ (\ref{Fried}-\ref{equ}) and taking also into account the current observational constraints and theoretical understanding, a list of the fundamental properties that an adiabatic UDM model has to satisfy in order to be viable. We then  translate these properties on the $p-\rho$ plane, see Fig.\ \ref{fig:p_rho}.

\begin{enumerate}
\item We assume the UDM to satisfy the weak energy condition: $\rho \geqslant 0$; therefore, we are only  interested in the positive half plane. In addition, we assume that the null energy condition is satisfied: $\rho+p\geq 0$, i.e.\ our UDM is a standard (non-phantom) fluid.
Finally, we assume that our UDM models admit a $\rho_{\Lambda}$, so that an asymptotic $w = -1$ is built in.

\item We demand a dust-like behaviour back in the past, at high energies,  i.e.\ a negligible pressure $p \ll \rho$ for $\rho \gg \rho_\Lambda$.\footnote{Note that we could have $p\simeq -\rho_{\Lambda}$ and yet, if $\rho \gg \rho_\Lambda$, the Universe would still be in a matter-like era.}  In particular, for an adiabatic fluid we require that at recombination $|w_{\rm rec}| \lesssim 10^{-6}$, see \cite{Muller:2004yb,Pietrobon:2008js,Balbi:2007mz,Quercellini:2007ht}.

\item Let us consider a Taylor expansion of the  UDM EoS  $p(\rho)$ about the present energy density $\rho_0$: 
\begin{equation}
\label{affine}
p \simeq p_0 + \alpha(\rho - \rho_0)\;,
\end{equation}
i.e.\ an ``affine'' EoS model \cite{Pietrobon:2008js,Balbi:2007mz,Quercellini:2007ht,Ananda:2005xp} where $\alpha$ is the  adiabatic speed of sound at the present time. Clearly, these models would be represented by straight lines in Fig.\  \ref{fig:p_rho}, with $\alpha$ the slope.
The $\Lambda$CDM model, interpreted as  UDM, corresponds to the affine model (\ref{affine}) with $\alpha = 0$ (see \cite{Ananda:2005xp} and \cite{Balbi:2007mz,Pietrobon:2008js}) and thus  it   is represented in Fig.\ \ref{fig:p_rho} by the horizontal line $p = -\rho_\Lambda$.
From the matter power spectrum constraints on affine models \cite{Pietrobon:2008js}, it turns out that $\alpha \lesssim 10^{-7}$. Note therefore that, from the UDM  perspective,  today we necessarily have $w\simeq -0.7$. 
\end{enumerate}

Few comments are in order. From the points above, one could conclude that any adiabatic UDM model, in order to be viable, necessarily has to degenerate into the $\Lambda$CDM model, as shown in \cite{Sandvik:2002jz} for the generalised Chaplygin gas and in \cite{Pietrobon:2008js} for the affine adiabatic model\footnote{From the point of view of the analysis of models in the $p-\rho$ plane of Fig.\ \ref{fig:p_rho}, the constraints found by Sandvik {\it et al} \cite{Sandvik:2002jz} on the generalised Chaplygin gas UDM models and by   \cite{Pietrobon:2008js}  on the affine UDM models simply amount to say that the curves representing these models are indistinguishable from the horizontal $\Lambda$CDM line.} (see \cite{Davis:2007na,Kessler:2009ys,Sollerman:2009yu} for an analysis of other models). In other words, one would conclude that any UDM model should satisfy the condition $c_{\rm s}^2 \ll 1$ at all times, so that   $k_{\rm J}^2 \gg k^2$ for all scales of cosmological interest, in turn giving an evolution for the gravitational potential $\Phi$ as  in Eq.\ (\ref{uklesskJ}):
\begin{equation}\label{kjggksol}
 \Phi_{\rm k} \simeq A_{\rm k}\left(1 - \frac{H}{a}\int a^2 d\eta\right)\;,
\end{equation}
where $A_{\rm k} = \Phi_{\rm k}\left(0\right)T_{\rm m}\left(k\right)$, $\Phi_{\rm k}\left(0\right)$ is the primordial gravitational potential at large scales, set during inflation,  and $T_{\rm m}\left(k\right)$ is the matter transfer function, see e.g.\ \cite{Dodelson:2003ft}.

On the other hand, let us write down the explicit form of the Jeans wave number:
\begin{equation}\label{kJ2analytic}
 k_{\rm J}^{2} = \frac{3}{2}\frac{\rho}{(1 + z)^2}\frac{(1 + w)}{c_{\rm s}^2}\left|\frac{1}{2}(c_{\rm s}^2 - w) - \rho\frac{dc_{\rm s}^2}{d\rho} + \frac{3(c_{\rm s}^2 - w)^2 - 2(c_{\rm s}^2 - w)}{6(1 + w)} + \frac{1}{3}\right|\;.
\end{equation}
Clearly, we can obtain  a large $k_{\rm J}^{2}$ not only when $c_{\rm s}^2 \to
0$, but also when $c_{\rm s}^2$ changes rapidly, i.e.\ when the
above expression is dominated by the $\rho\; dc_{\rm s}^2/d\rho$
term. When this term is dominating in Eq.\ (\ref{kJ2analytic}), we may say that the EoS is characterised by a fast transition.

Thus, viable adiabatic UDM models can be constructed which do not require $c_{\rm s}^2 \ll 1$  at all times if the speed of sound goes through a rapid change, a fast transition period during which  $k_{\rm J}^{2}$ can remain large, in the sense that  $k^{2} \ll k_{\rm J}^{2}$ for all scales of cosmological interest to which the linear perturbation theory of Eq.\ (\ref{equ}) applies.  From point 3 above, at late times we must have $p \simeq - \rho_\Lambda$; on the other hand, at  recombination we have  $|w_{\rm rec}| \lesssim 10^{-6}$ and the speed of sound is negligible, implying \ $p \approx {\rm constant}$. Therefore, the transition will mark the passage from a very small (possibly vanishing) almost constant $p$ to the asymptotic value  $p \simeq - \rho_\Lambda$ or, in other words, from a pure CDM-like early phase to a post-transition $\Lambda$CDM-like late epoch. In addition, we may expect the transition to occur at relatively high redshifts, high enough to make the UDM model quite similar to the $\Lambda$CDM model at late times. Indeed,  again from point 3 above, we infer that the fast transition should take place sufficiently far in the past, in particular during the dark matter epoch, when $\rho \gg \rho_\Lambda$. Otherwise, we expect that it would be problematic to reproduce the current observations related to the UDM parameter $w$, for instance it would be hard to have a  good fit of supernovae  and ISW effect data. 

In the rest of the paper, in order to quantitatively investigate observational constraints  on UDM models with fast transition, we introduce and discuss a  toy model. In particular, we will  explore which values of the parameters of this toy model fit the observed CMB and matter power spectra. 

Finally, let us make a last remark on building  phenomenological UDM (or DE) fluid models intended to represent the homogeneous FLRW background and its linear perturbations. A fast transition in a fluid model could be characterised by a large value of $c_{\rm s}^{2}$, even larger than $1$. As far as the FLRW background evolution is concerned, this fact does not raise any issue: the background is homogeneous, and $c_{\rm s}^{2} = \dot{p}/\dot{\rho}$ does not actually represent a speed of sound, as there is nothing that could propagate in this case. For linear perturbations, at scales such that $k \ll k_{\rm J}$ the solution of the Eq.\ (\ref{equ}) for the gravitational potential is Eq.\ (\ref{kjggksol}). Therefore, for such scales there is no superluminal propagation. This is because Eq.\ (\ref{equ}) is the Fourier component of a wave equation with potential ${\theta''}/{\theta}$, and this potential does not allow propagation for $k \ll k_{\rm J}$. In building a phenomenological fluid model, we can therefore choose values for the parameters of the model in order to always satisfy the condition  $k \ll k_{\rm J}$ for all $k$ of cosmological interests to which linear theory applies, hence such a fluid model will be a good causal model for all scales that intends to represent. In other words, we can always build the fluid model in such a way that  all scales smaller than the Jeans length $\lambda \ll \lambda_{\rm J}$
 correspond to those in the non-linear regime, i.e.\ scales beyond the range of applicability of the model.
So, for these scales, no conclusions can be derived from the linear theory on the behaviour of perturbations of a UDM model with $c_{\rm s}^{2} \gtrsim 1$.  To investigate these scales, one needs to be beyond the perturbative regime investigated here, possibly also increasing the sophistication  of the fluid model in order to properly take into account  the greater complexity of small scale non-linear physics and to maintain causality.

\section{A toy model with fast transition}\label{sec:tghmodel}

\begin{figure}[htbp]
\begin{center}
\includegraphics[width=\columnwidth]{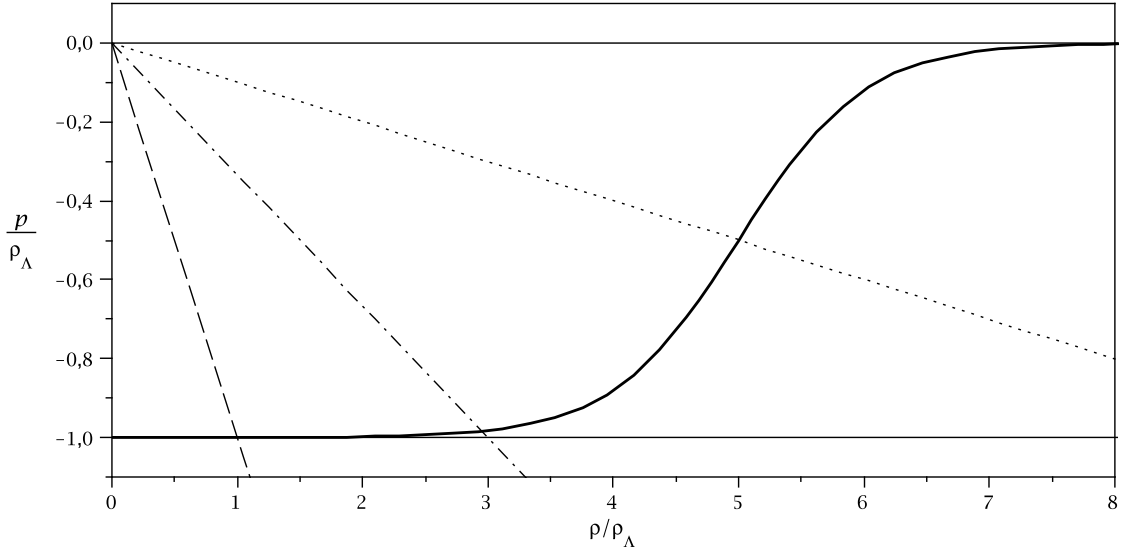}
\caption{Illustrative plot of the EoS as a functions of the energy density for the hyperbolic tangent model. The  parameters values are $\rho_{\rm t}/\rho_\Lambda = 5$ and $\rho_{\rm s}/\rho_\Lambda = 1$. The energy density and the pressure are normalised to $\rho_\Lambda$. The other five lines are the ones plotted and described in Fig.\ 1.}
\label{Fig2eos}
\end{center}
\end{figure}

In the present section we introduce a toy model based on a hyperbolic tangent EoS, which is conveniently parametrised as
\begin{equation}\label{tanh}
p = -\rho_\Lambda\left[\frac{1 - \tanh\left(\frac{\rho - \rho_{\rm t}}{\rho_{\rm s}}\right)}{1 - \tanh\left(\frac{\rho_\Lambda - \rho_{\rm t}}{\rho_{\rm s}}\right)}\right]\;,
\end{equation}
and is depicted in Fig.\ \ref{Fig2eos} for a typical shape. In the EoS (\ref{tanh}) $\rho_{\rm t}$ is the typical energy scale at the transition, $\rho_{\rm s}$ is related to the rapidity of the transition, $\rho_\Lambda$ is the effective cosmological constant, i.e.\ $p(\rho_{\Lambda})=-\rho_{\Lambda}$. This model reduces to a $\Lambda$CDM, which in the UDM language of the previous section is described by $p = -\rho_{\Lambda}$, in two limits: \\ $\rho_{\rm t} \to \infty$ and $\rho_{\rm s}\to \infty$. \\

The main properties of the EoS (\ref{tanh}) are the following:
\begin{enumerate}
 \item The asymptotic behaviour of the pressure for $\rho \gg \rho_{\rm t}$ is  $p \approx 0$. From the considerations of the previous section, we expect $\rho_{\rm t} \gg \rho_\Lambda$, which corresponds to a minimum value  of the redshift $z_{\rm t}$. For instance, we have $z_{\rm t} \gtrsim 1.85$ if we want to have  $\rho_{\rm t} \gtrsim 10\rho_\Lambda$.
In Figs.\ \ref{wtanhzFig2} we plot the evolution of $w$ as a function of  redshift, for $\rho_{\rm t}/\rho_\Lambda = 10$ (left panel) and $\rho_{\rm t}/\rho_\Lambda = 20$ (right panel), for three different choices of $\rho_{\rm s}/\rho_\Lambda$. The solid line represents the $\Lambda$CDM model, while the horizontal lines respectively represent:   a pure CDM model for  $w=0$; the boundary between the decelerated  and the accelerated expansion phases of the Universe for  $w = -1/3$. 
Clearly, from both panels, models with larger $\rho_{\rm s} /\rho_{\rm t}$ ratio have a background evolution  more similar to that of the $\Lambda$CDM model at all times.
On the other hand, a smaller  $\rho_{\rm s}/\rho_{\rm t}$ ratio implies a faster transition between the CDM-like phase and the $\Lambda$CDM phase. In addition, we have the confirmation that the  transition has to take place sufficiently far in the past, i.e.\ $\rho_{\rm t} \gg \rho_\Lambda$, in order for  the late time evolution of $w$ to be  in any case close to that of the  $\Lambda$CDM model.
\begin{figure}[htbp]
\begin{center}
\includegraphics[width=0.49\columnwidth]{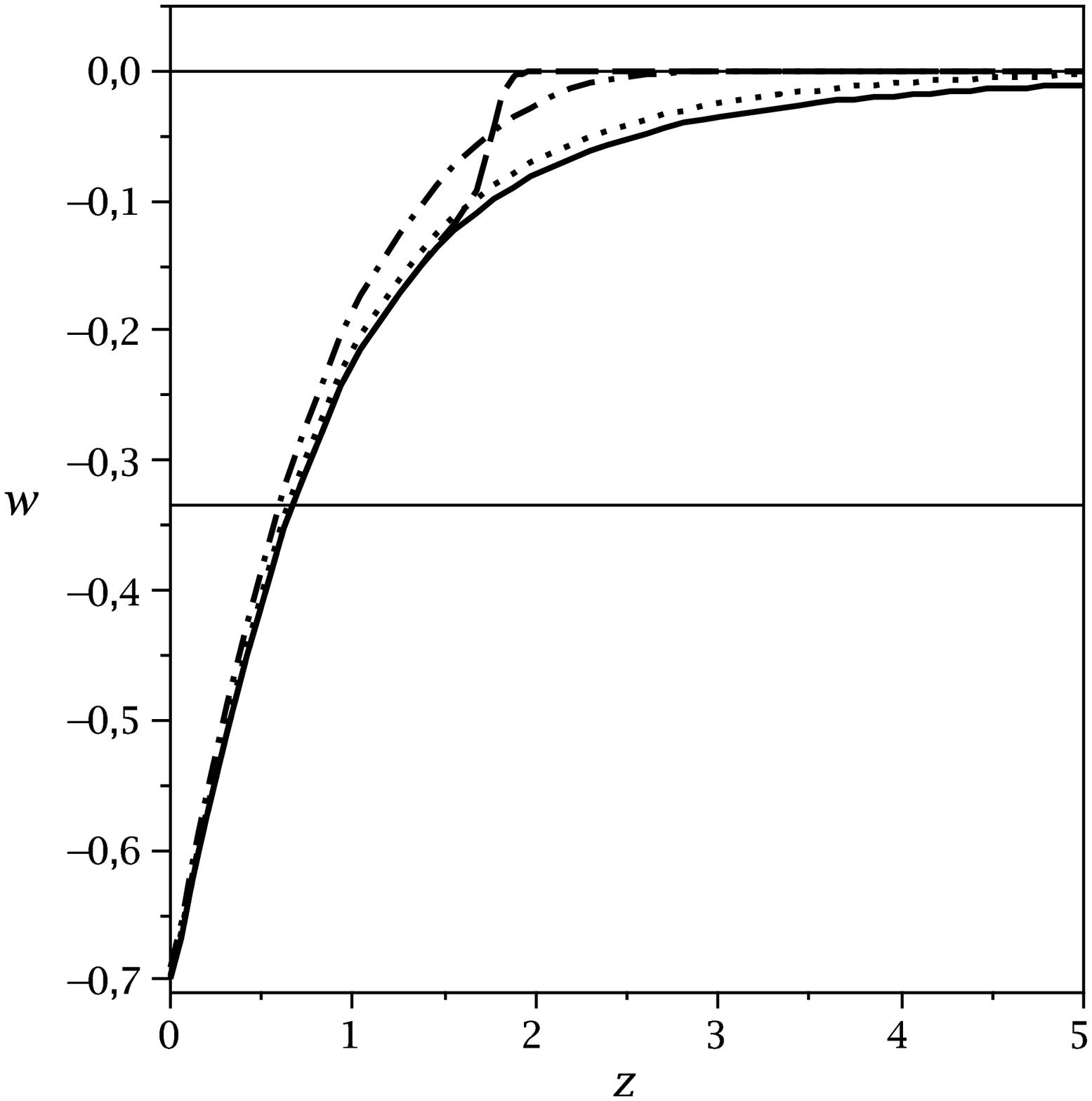}
\includegraphics[width=0.49\columnwidth]{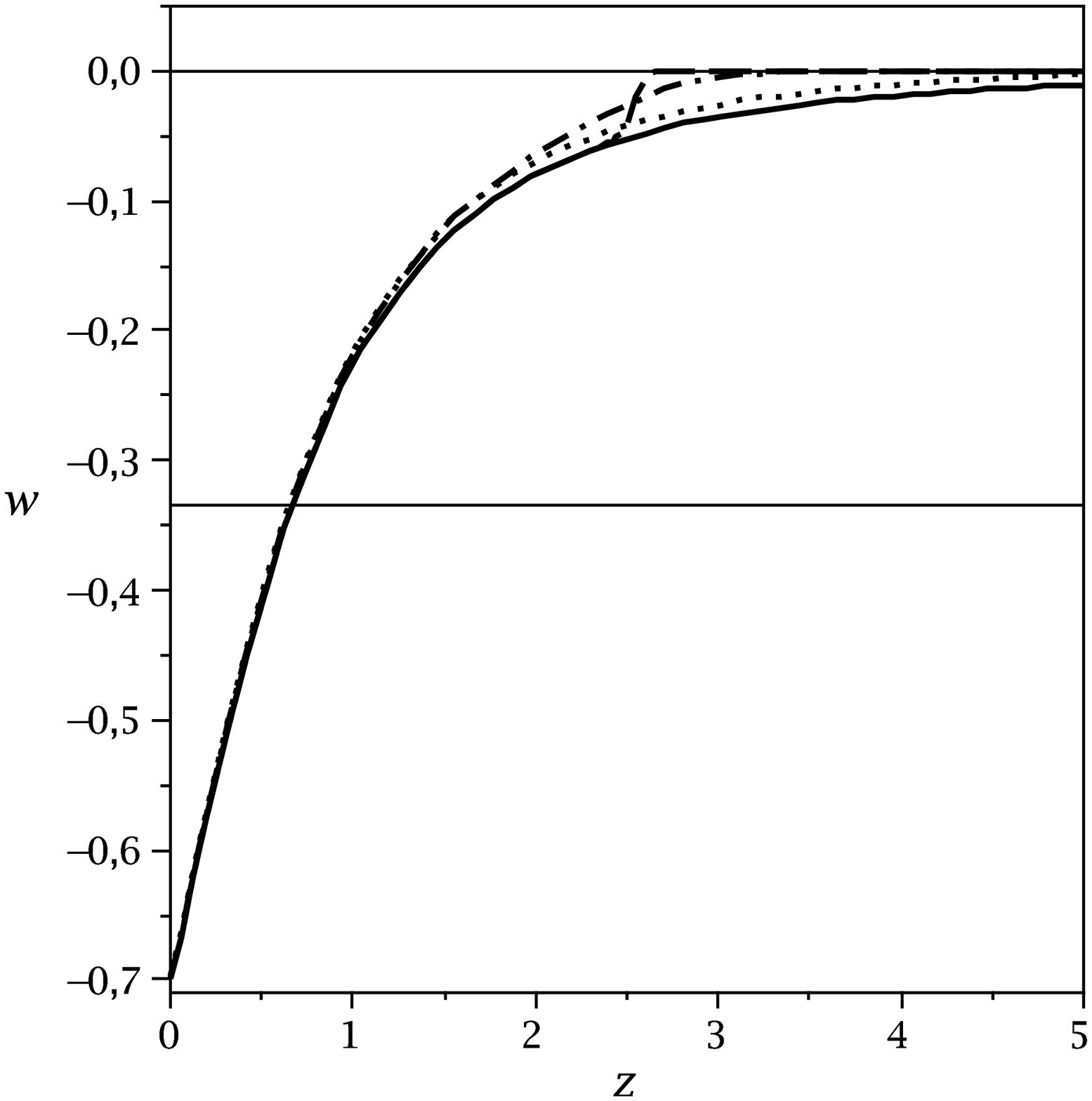}
\caption{Evolution of the UDM parameter $w = p/\rho$ in the hyperbolic tangent model, for $\rho_{\rm t}/\rho_\Lambda = 10$ (left panel) and $\rho_{\rm t}/\rho_\Lambda = 20$ (right panel). For reference we also plot: the  $w = 0$ line representing a   flat pure CDM model (an EdS Universe); the  $w=-1/3$ line representing the boundary between the decelerated and the accelerated  phases; the solid curve representing the evolution of $w$ for a typical $\Lambda$CDM model with $\Omega_{\Lambda} = 0.7$. In each panel, the three dashed, dash-dotted and dotted lines respectively correspond to $\rho_{\rm s}/\rho_\Lambda = 1, 10, 100$. Clearly, the dotted lines correspond to  UDM models with a slow transition, almost indistinguishable from a $\Lambda$CDM at all times, while the dashed lines well represent  UDM models with a very fast transition from a pure CDM to a typical  $\Lambda$CDM behaviour. The higher  $\rho_{\rm t}/\rho_\Lambda $, the earlier the UDM $w$ transits to that of a $\Lambda$CDM.}
\label{wtanhzFig2}
\end{center}
\end{figure}

\begin{figure}[htbp]
\begin{center}
\includegraphics[width=0.5\columnwidth]{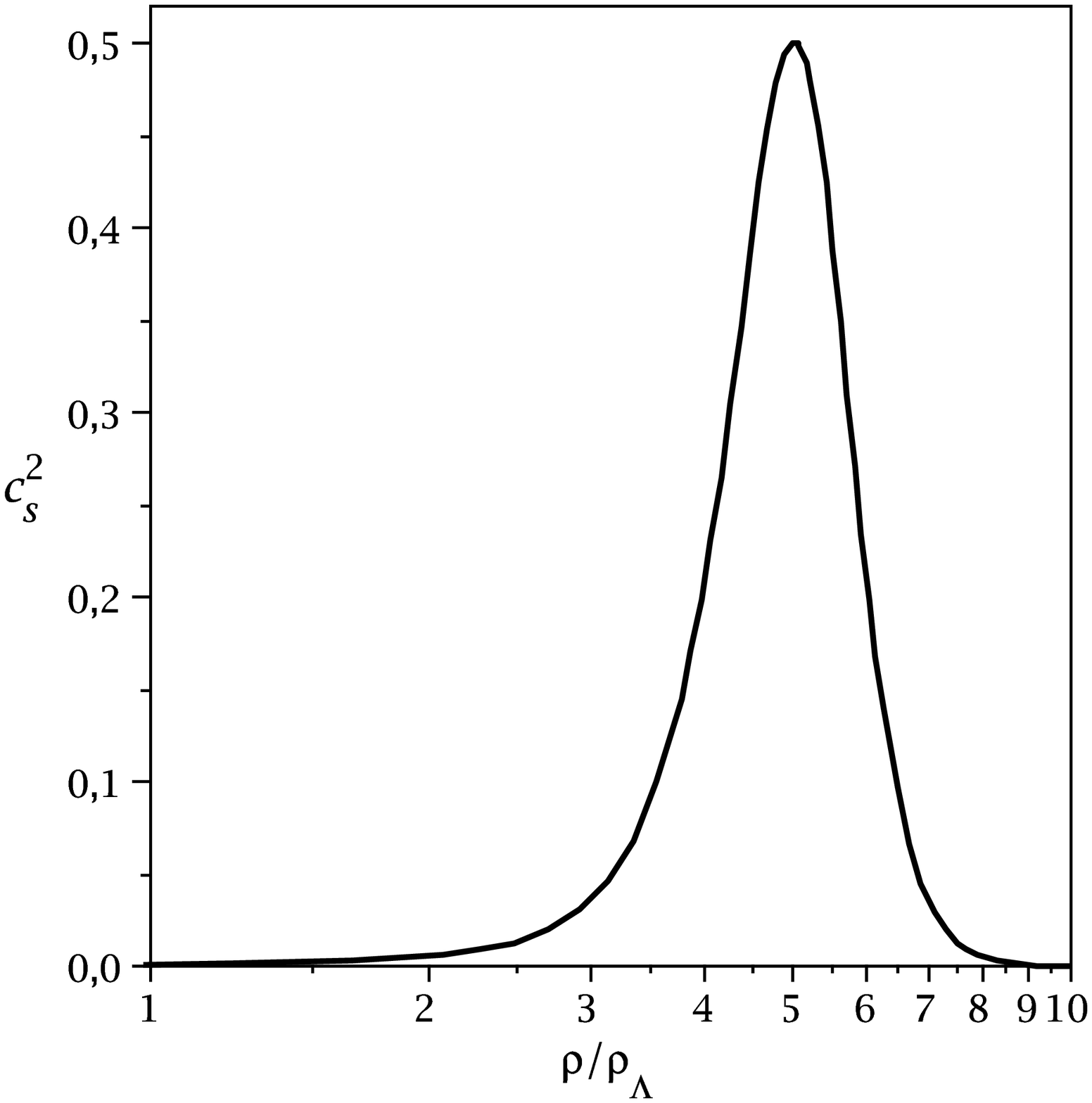}\includegraphics[width=0.5\columnwidth]{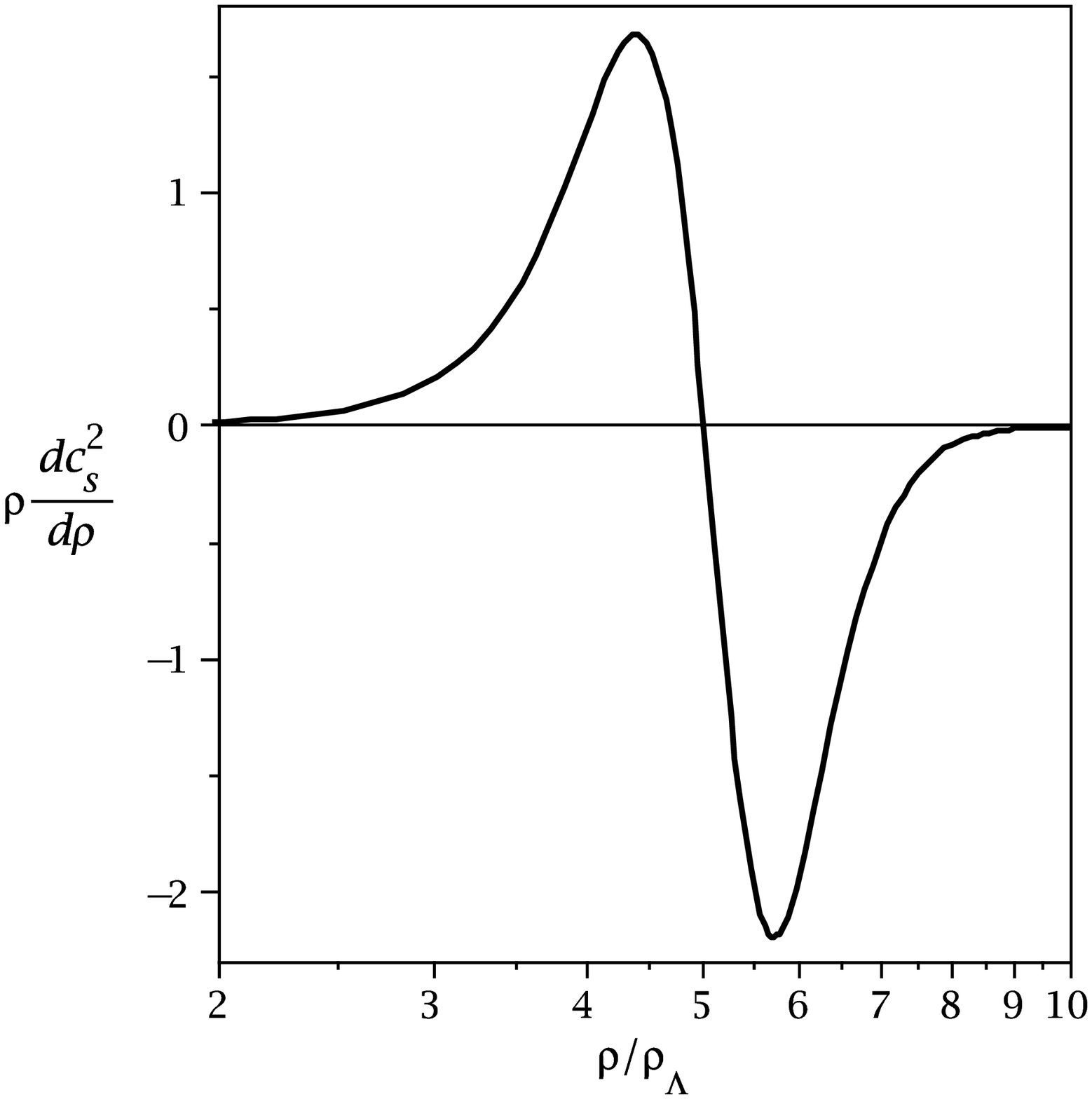}
\caption{Illustrative plots of  the speed of sound and $\rho\; dc_{\rm s}^2/d\rho$ as  functions of the energy density for the hyperbolic tangent model. The  parameters values are $\rho_{\rm t}/\rho_\Lambda = 5$ and $\rho_{\rm s}/\rho_\Lambda = 1$. The energy density and the pressure are normalised to $\rho_\Lambda$.}
\label{FigSoS}
\end{center}
\end{figure}

\item The speed of sound is the following:
\begin{equation}\label{tanhcs2}
 c_{\rm s}^{2} = \frac{\rho_\Lambda}{\rho_{\rm s}}\frac{1 - \tanh^{2}\left(\frac{\rho - \rho_{\rm t}}{\rho_{\rm s}}\right)}{1 - \tanh\left(\frac{\rho_\Lambda - \rho_{\rm t}}{\rho_{\rm s}}\right)}\;,
\end{equation}
illustrated in the  left panel of Fig.\ \ref{FigSoS}.
It attains its maximum value
\begin{equation}
\label{csmax}
 \left.c_{\rm s}^2\right|_{\rm max}=\frac{\rho_\Lambda/\rho_{\rm s}}{1-\tanh\left(\frac{\rho_\Lambda-\rho_{\rm t}}{\rho_{\rm s}}\right)}
\end{equation}
 in $\rho = \rho_{\rm t}$. For our analysis, there are two main cases to consider, assuming $\rho_{\rm t}\gg \rho_{\Lambda}$:
\begin{enumerate}
\item[{\it a)} ] $\rho_{\rm t} \ll \rho_{\rm s}$. In this case, $\left.c_{\rm s}^2\right|_{\rm max} \sim \rho_\Lambda/\rho_{\rm s} \sim 0$, so that the model is close to a $\Lambda$CDM at all times.
\item[{\it b)} ] $\rho_{\rm t} \gg \rho_{\rm s}$.  In this case, we have two subcases: {\it bi)}   $\rho_\Lambda \lesssim 2 \rho_{\rm s}$, for which $\left.c_{\rm s}^2\right|_{\rm max} \sim \rho_\Lambda/2\rho_{\rm s} < 1$ or {\it bii)}  $  \rho_\Lambda \gtrsim 2\rho_{\rm s}$, for which $\left.c_{\rm s}^2\right|_{\rm max} \sim \rho_\Lambda/2\rho_{\rm s} > 1$. The latter subcase may in principle imply  superluminal perturbations; fortunately, as we shall see, a-causal effects can be avoided  if the transition is sufficiently fast.
\end{enumerate}

\item As we explained in the previous section, in order to have a fast transition we must have  $\rho\, dc_{\rm s}^2/d\rho\gg 1$ in Eq.\ (\ref{kJ2analytic}) for the  Jeans wave number. This quantity is depicted in the right panel of Fig.\ \ref{FigSoS}. For the EoS (\ref{tanh}) the derivative of $c_{\rm s}^{2}$ is
\begin{equation}
 \frac{dc_{\rm s}^{2}}{d\rho} = -2\frac{\rho_\Lambda}{\rho_{\rm s}^2}\tanh\left(\frac{\rho - \rho_{\rm t}}{\rho_{\rm s}}\right)\frac{1 - \tanh^{2}\left(\frac{\rho - \rho_{\rm t}}{\rho_{\rm s}}\right)}{1 - \tanh\left(\frac{\rho_\Lambda - \rho_{\rm t}}{\rho_{\rm s}}\right)} = -\frac{2}{\rho_{\rm s}}\tanh\left(\frac{\rho - \rho_{\rm t}}{\rho_{\rm s}}\right)c_{\rm s}^{2}\;,
\end{equation}
which attains its extrema at $\rho = \rho_{\rm t} \pm \rho_{\rm s}\tanh^{-1}\left(\sqrt{3}/3\right) \simeq \rho_{\rm t} \pm 0.66\rho_{\rm s}$. The maximum corresponds to the minus sign.

Clearly, the derivative of $c_{\rm s}^{2}$ is important only in the case {\it b)} of the previous point. In this case the maximum is:
\begin{equation}\label{dcs2drhomax}
 \left.\rho\frac{dc_{\rm s}^{2}}{d\rho}\right|_{\rm max} \simeq \left.\rho_{\rm t}\frac{dc_{\rm s}^{2}}{d\rho}\right|_{\rm max} \simeq \frac{\rho_\Lambda}{\rho_{\rm s}}\frac{\rho_{\rm t}}{\rho_{\rm s}}\;.
\end{equation}
For subcase {\it bii)}, $c_{\rm s}^{2}>1$, we always have $\rho_{\Lambda}\rho_{\rm t}/\rho_{\rm s}^2 \gg 1$, while in subcase {\it bi)} there is also the possibility that $\rho_{\Lambda}\rho_{\rm t}/\rho_{\rm s}^2$ be small.

Let us now consider the case when the transition takes place at the lower limit $z_{\rm t} \sim 1.85$, corresponding to $\rho_{\rm t} \sim 10 \rho_\Lambda$. In this case, from Eq.\  (\ref{dcs2drhomax}), the maximum is $10\rho_\Lambda^2/\rho_{\rm s}^2$. Therefore, in order to have a fast transition, we must have $\rho_\Lambda \gtrsim \rho_{\rm s}$. Then, it is inevitable from point {\it bii)} that, if we want the fast transition to take place just before the accelerated phase of the expansion of the Universe, we must have $c_{\rm s}^{2} > 1$. In this case, shortly  after the transition the pressure rapidly approaches  the asymptotic value $-\rho_\Lambda$.
\end{enumerate}

\section{Analysis of the Jeans wave number during the transition}\label{sec:perts}

The Jeans length is a crucial quantity in determining the viability of a UDM model, because of its effect on perturbations, which is then revealed in observables such as the CMB and matter power spectra. We now focus on the Jeans wave number for the toy UDM model introduced in the previous section and investigate its behaviour as a function of the speed of sound, in particular around $\rho = \rho_{\rm t}$, in the middle of the transition where the speed of sound  is at its peak.

Starting from the classification we presented in point 2 of the previous section, we are interested in the case {\it b)}, namely $\rho_{\rm t} \gg \rho_{\rm s}$, because in this regime a fast transition in the EoS takes place.
The majority of the adiabatic UDM models considered so far in the literature belongs to the case {\it a)} $\rho_{\rm t} \ll \rho_{\rm s}$ of  point 2 of section \ref{sec:tghmodel}. For the toy model Eq.\ (\ref{tanh}) as well, $\rho_{\rm t} \ll \rho_{\rm s}$ implies that the pressure tends to $p \simeq -\rho_\Lambda$ at all times, i.e.\ to a $\Lambda$CDM, as shown in Fig.\ \ref{wtanhzFig2}.

\begin{figure}[htbp]
\includegraphics[width=0.5\columnwidth]{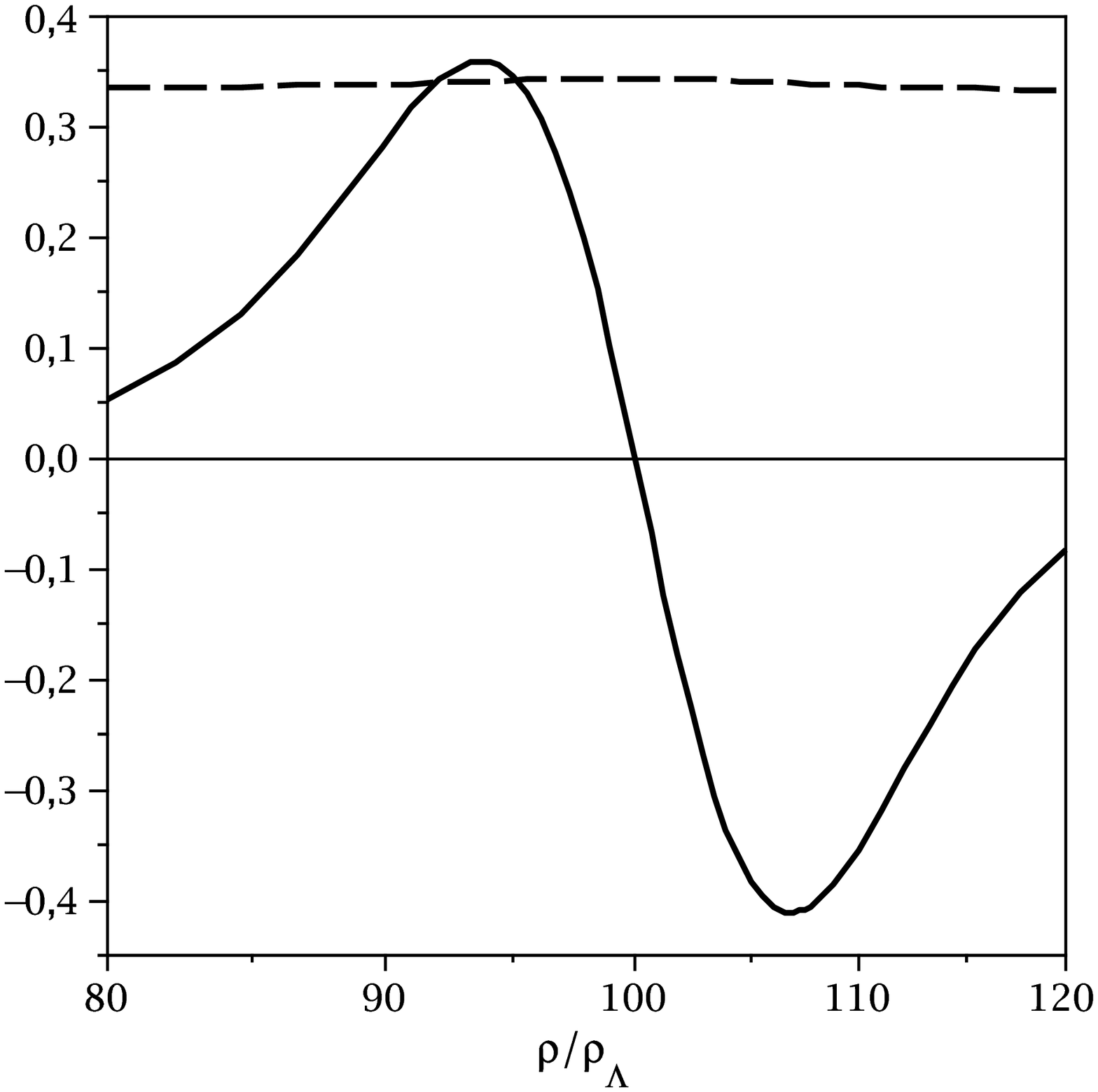}\includegraphics[width=0.5\columnwidth]{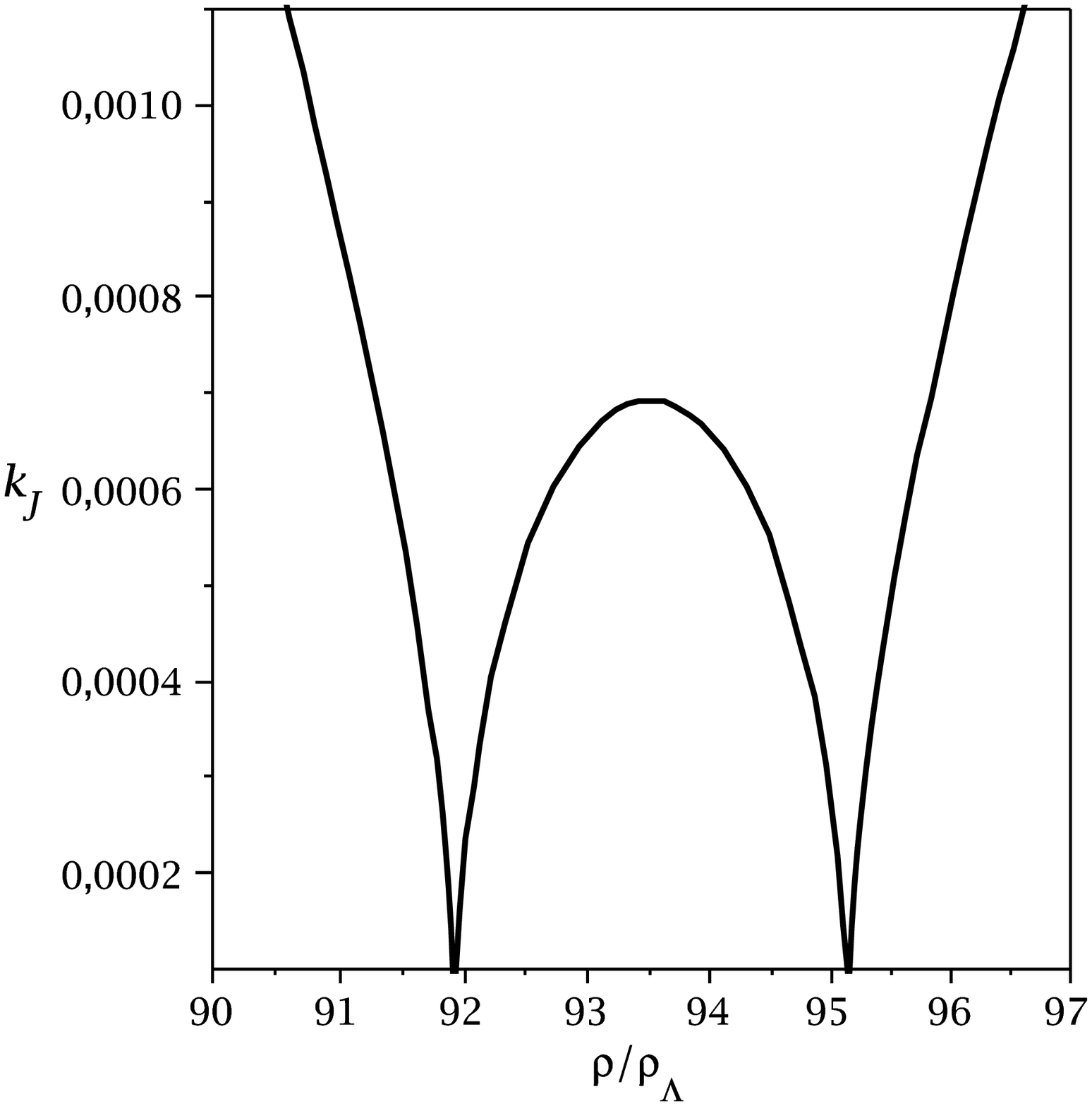}
\caption{Left panel: evolution of $\rho\,{ d} c_{\rm s}^2/{d}\rho$ (solid line) and $\mathcal{B}$ (dashed line) as functions of the energy density. Right panel: evolution of the Jeans wave number as function of the energy density $\rho/\rho_\Lambda$. The Jeans wave number $k_{\rm J}$ is in $h$ Mpc$^{-1}$ units and it has been calculated assuming $\Omega_\Lambda = 0.7$. The choice of the parameters is: $\rho_{\rm t} = 100\rho_\Lambda$ ($z_t\simeq 5.14$) and $\rho_{\rm s} = 10\rho_\Lambda$. }
\label{comparison3}
\end{figure}

\begin{figure}[htbp]
\includegraphics[width=0.5\columnwidth]{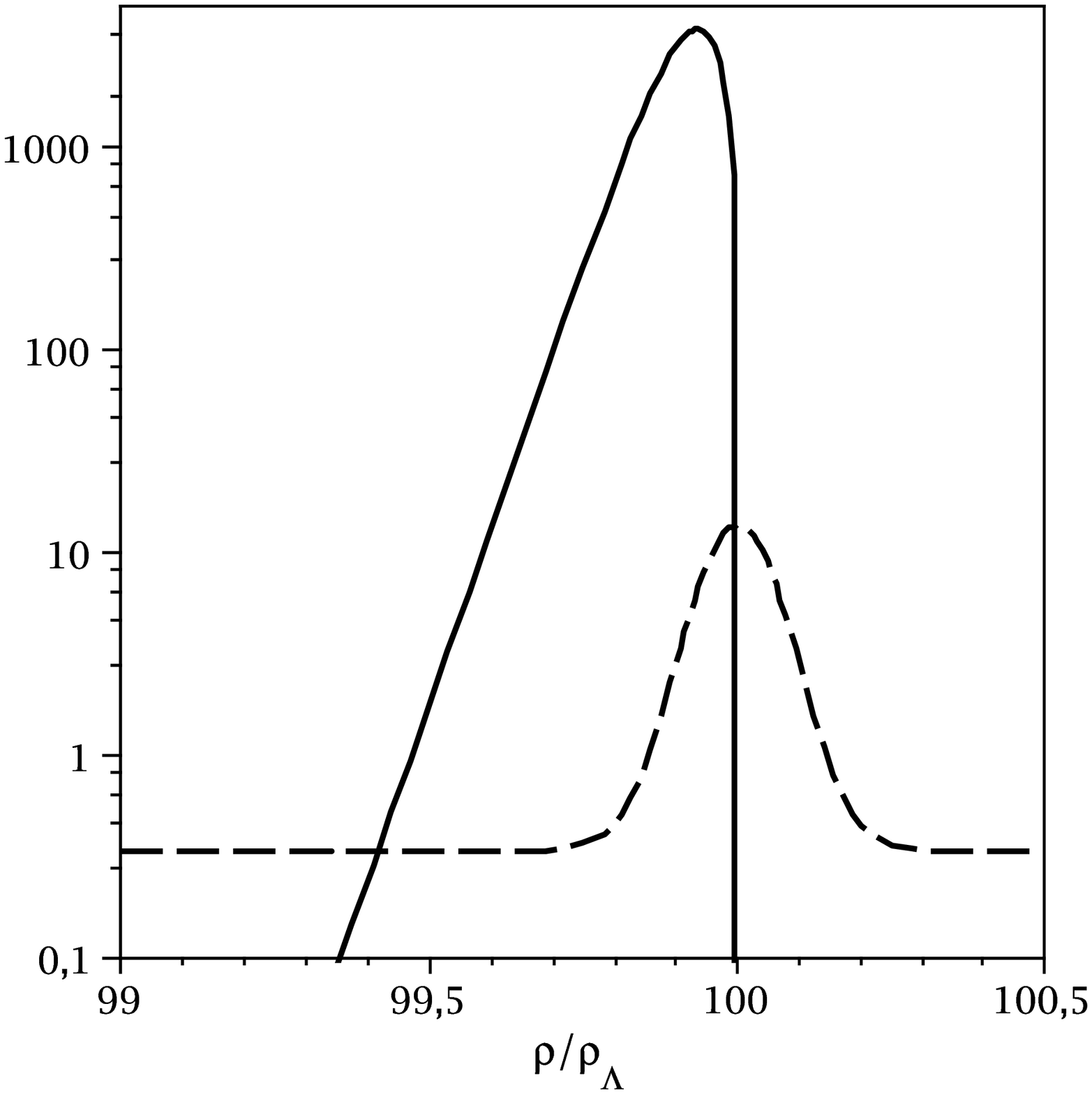}\includegraphics[width=0.5\columnwidth]{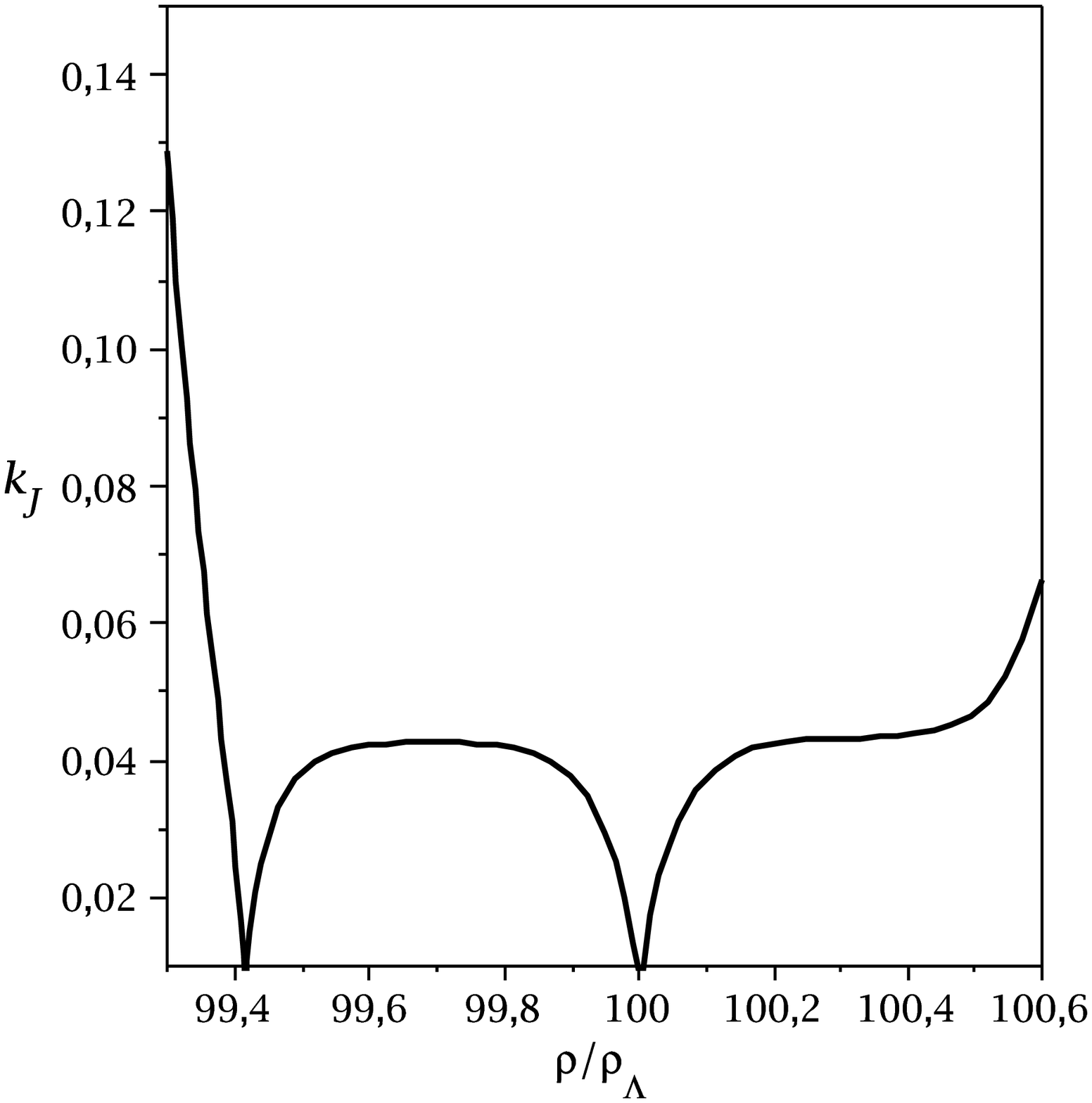}
\caption{Same as in Fig.\ 4, again with    $\rho_{\rm t} = 100\rho_\Lambda$ ($z_t\simeq 5.14$)
 but now with  $\rho_{\rm s} = 10^{-1}\rho_\Lambda$.}
\label{comparison2}
\end{figure}

\begin{figure}[htbp]
\includegraphics[width=0.5\columnwidth]{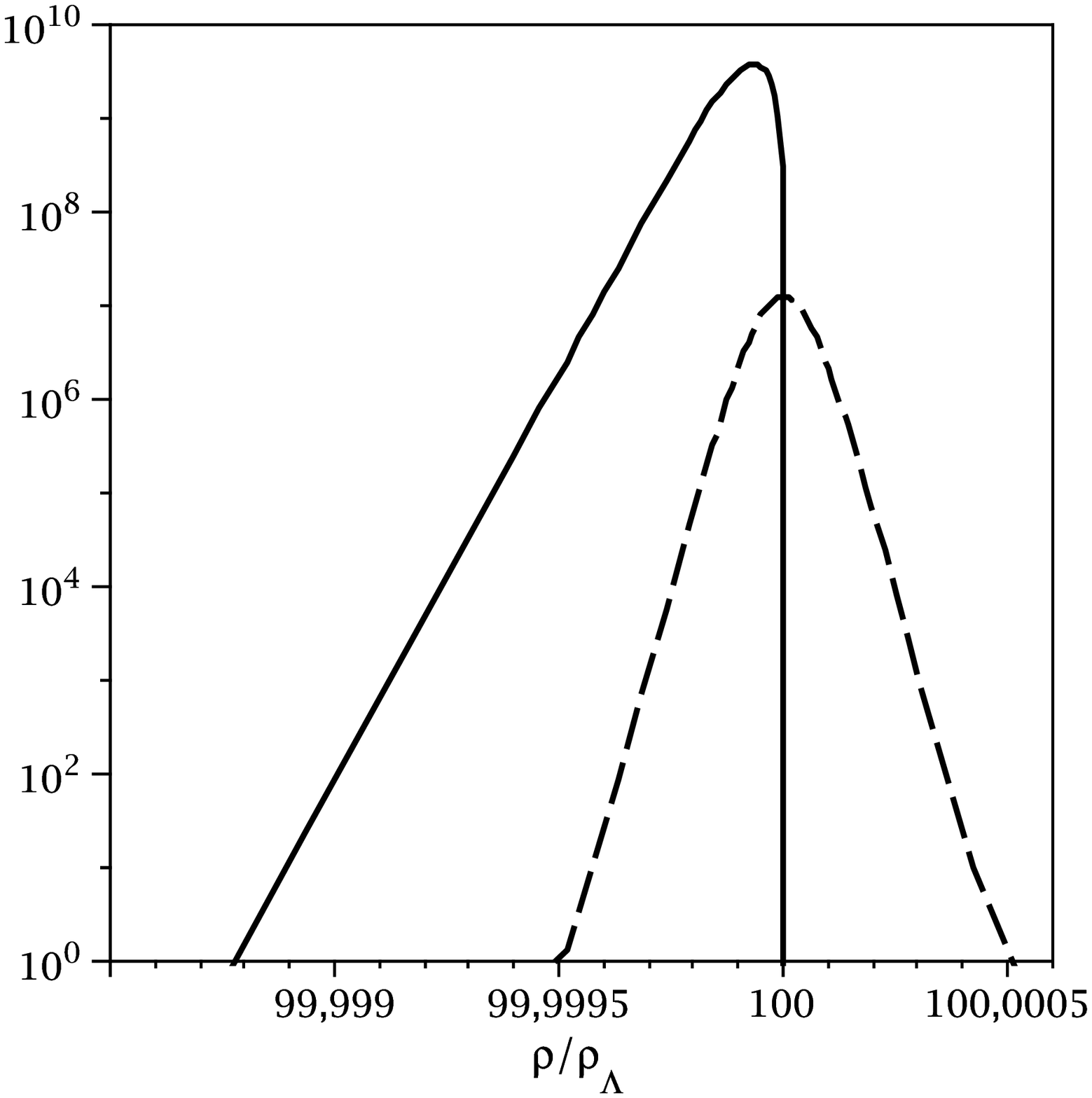}\includegraphics[width=0.5\columnwidth]{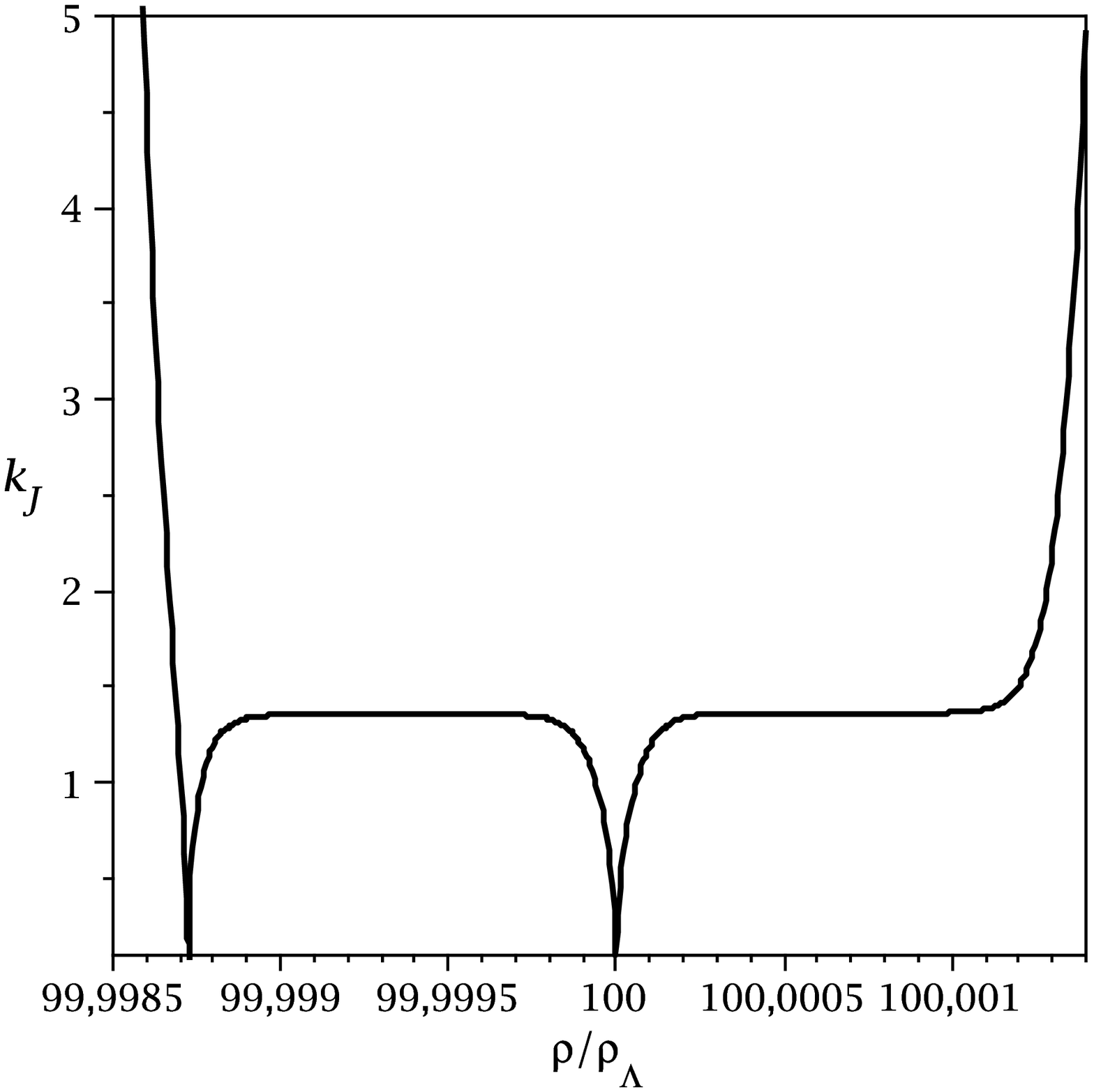}
\caption{Same as in Figs.\ 4-5,  again with   $\rho_{\rm t} = 100\rho_\Lambda$ ($z_t\simeq 5.14$) but now with  $\rho_{\rm s} = 10^{-4}\rho_\Lambda$.}
\label{comparison1}
\end{figure}

In the case of a fast transition, from Eq.\ (\ref{kJ2analytic}) for the Jeans wave number, it is interesting to compare the term $\rho\,{d}c_{\rm s}^2/{d}\rho$ with the remaining ones contained in the squared brackets, namely:
\begin{equation}
 \mathcal{B}:= \frac{1}{2}(c_{\rm s}^2 - w) + \frac{3(c_{\rm s}^2 - w)^2 - 2(c_{\rm s}^2 - w)}{6(1 + w)} + \frac{1}{3}\;.
\end{equation}
In Figs.\ \ref{comparison3}-\ref{comparison1} we plot $\rho\,{d} c_{\rm s}^2/{ d}\rho$, $\mathcal{B}$ and the Jeans wave number $k_{\rm J}$ as functions of $\rho/\rho_\Lambda$. For the calculation of $k_{\rm J}$ we use $\rho_\Lambda = \Omega_\Lambda\rho_0$, with $\Omega_\Lambda = 0.7$ and the critical energy density $\rho_0 = 3H_0^2$, where $H_0$ is the Hubble constant. We choose  $\rho_{\rm t} = 100\rho_\Lambda$ in order to consider a transition sufficiently back in the Dark Matter epoch (see point 1 of section \ref{sec:tghmodel}) and vary the ratio $\rho_{\rm s}/\rho_\Lambda$, with  $\rho_{\rm s} = 10\rho_\Lambda, 10^{-1}\rho_\Lambda, 10^{-4}\rho_\Lambda$, in order to show examples of faster transitions.
From the plots in Figs.\ \ref{comparison3}-\ref{comparison1} it is clear that the smaller $\rho_{\rm s}/\rho_{\Lambda}$ is, the larger is the difference between $\rho\, {d} c_{\rm s}^2/{ d}\rho$ and $\mathcal{B}$. 

Moreover, from Fig.\ \ref{comparison3}, $\rho\, {d} c_{\rm s}^2/{ d}\rho$ is negative for $\rho > \rho_{\rm t}$, then for $\rho \approx \rho_{\rm t}$ it increases becoming positive and intersecting the $\mathcal{B}$ curve a first time for $\rho < \rho_{\rm t}$. For smaller values of the energy density, $\rho\, { d} c_{\rm s}^2/{ d}\rho$ decreases again to zero, again  intersecting  the $\mathcal{B}$ curve. In Figs.\ \ref{comparison2}-\ref{comparison1}, the same behaviour of the curves takes place and since the difference between the two curves is much larger, we have chosen a logarithmic scale. Therefore, the negative part of $\rho\, { d} c_{\rm s}^2/{ d}\rho$ has been omitted.

The intersection points between the curves $\rho\, { d} c_{\rm s}^2/{d}\rho$ and $\mathcal{B}$ represent the moments  at which the Jeans wave number $k_{\rm J}$ vanishes, as it can be seen from the right panels of Figs.\ \ref{comparison3}-\ref{comparison1}. In general, around these points the corresponding Jeans length becomes very large, possibly causing all sort of problems to perturbations, with effects on CMB and structure formation in the UDM model. On the other hand, for sufficiently small $\rho_{\rm s}$ the transition is fast enough that {\it i)} in general the Jeans wave number becomes larger and {\it ii)} it becomes vanishingly small for extremely short times, so that the the effects caused by its  vanishing  are sufficiently negligible,  as we are going to show in the next section for the CMB and matter power spectra.
As illustrated in  Figs.\ \ref{comparison3}-\ref{comparison1}, by choosing progressively smaller values of $\rho_{\rm s}$ we can obtain progressively larger Jeans wave numbers, while the curve $k_{\rm J}(\rho)$  starts to show a plateau shape around the transition. 

Clearly, we are interested in the value of $k_{\rm J}$ during the transition, because before and after that the negligible speed of sound implies a vanishing Jeans length, or a very large $k_{\rm J} $.
In essence, for a fast enough transition the ``average'' value of $k_{\rm J}$ around the transition is approximated by its value on the plateau - say $\hat{k}_{\rm J}$ - and this is, on average, the minimum value of $k_{\rm J}$, i.e.\ the maximum Jeans length for the given values of the parameters $\rho_{\rm s}$ and $\rho_{\rm t}$. Thus, we now want to establish a relation between $\rho_{\rm s}$ and $\rho_{\rm t}$ for any given $\hat{k}_{\rm J}$. This,  fixing  a $\hat{k}_{\rm J}$ which allows for an acceptable matter power spectrum which fits observational data, will help us to find the $\rho_{\rm s}$ needed to have the transition at $\rho_{\rm t}$. 

The relative  maximum of the Jeans wave number between the two zeros of the curve $k_{\rm J}(\rho)$  corresponds approximatively to where ${ d} c_{\rm s}^2/{ d}\rho$ assumes its maximum value, i.e.\ in $\hat{\rho} \simeq \rho_{\rm t} - 0.66\rho_{\rm s}$, as we have shown in point 3 of section \ref{sec:tghmodel}. Thus, let us define with $\hat{k}_{\rm J}$ the value of  ${k}_{\rm J}$ for $\rho = \hat{\rho}$: as required, it is of the same order of the plateau value (see for example Fig.\ \ref{comparison1}) of the Jeans wave number during the fast transition.

\begin{figure}[htbp]
\begin{center}
\includegraphics[width=0.5\columnwidth]{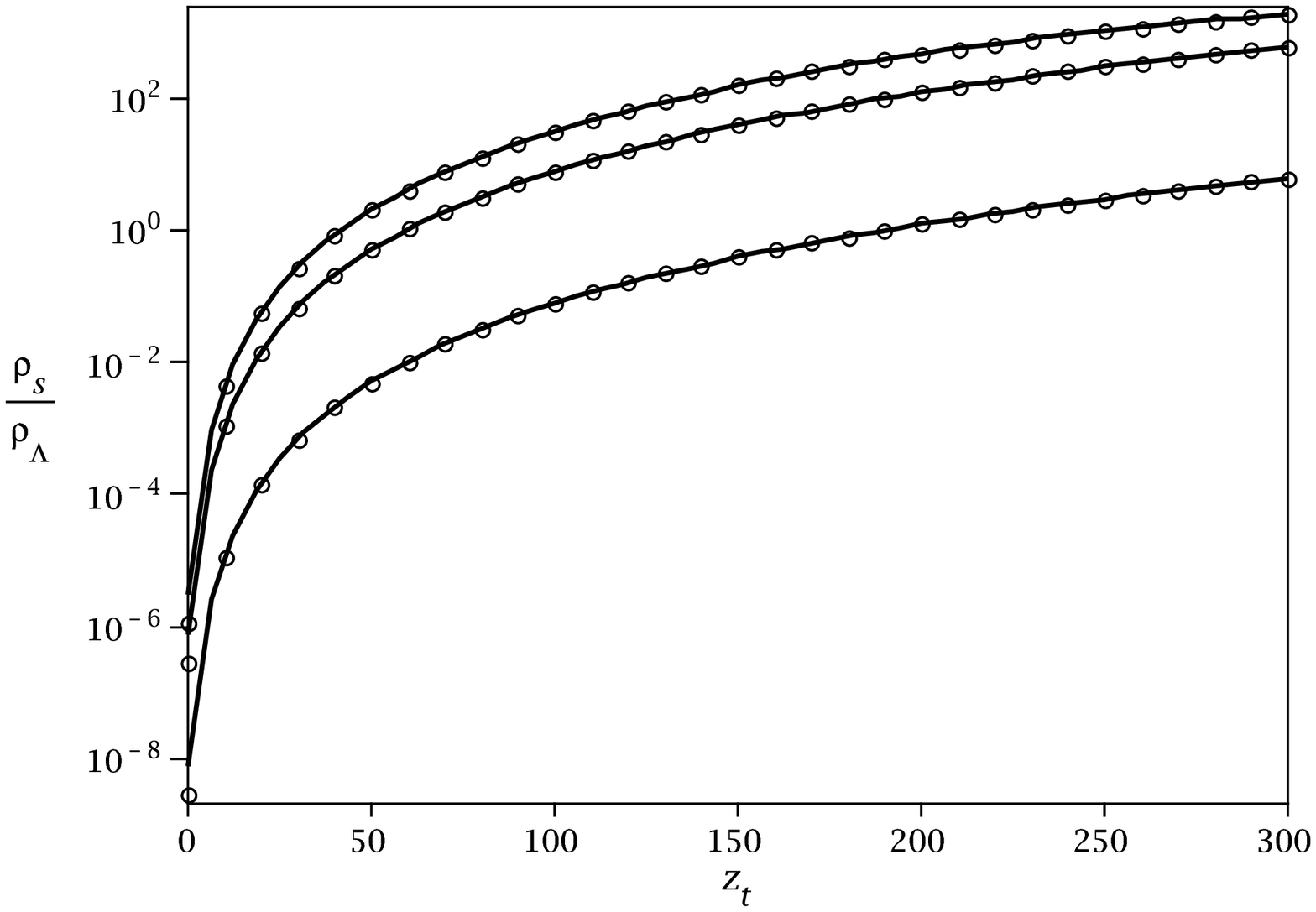}\includegraphics[width=0.5\columnwidth]{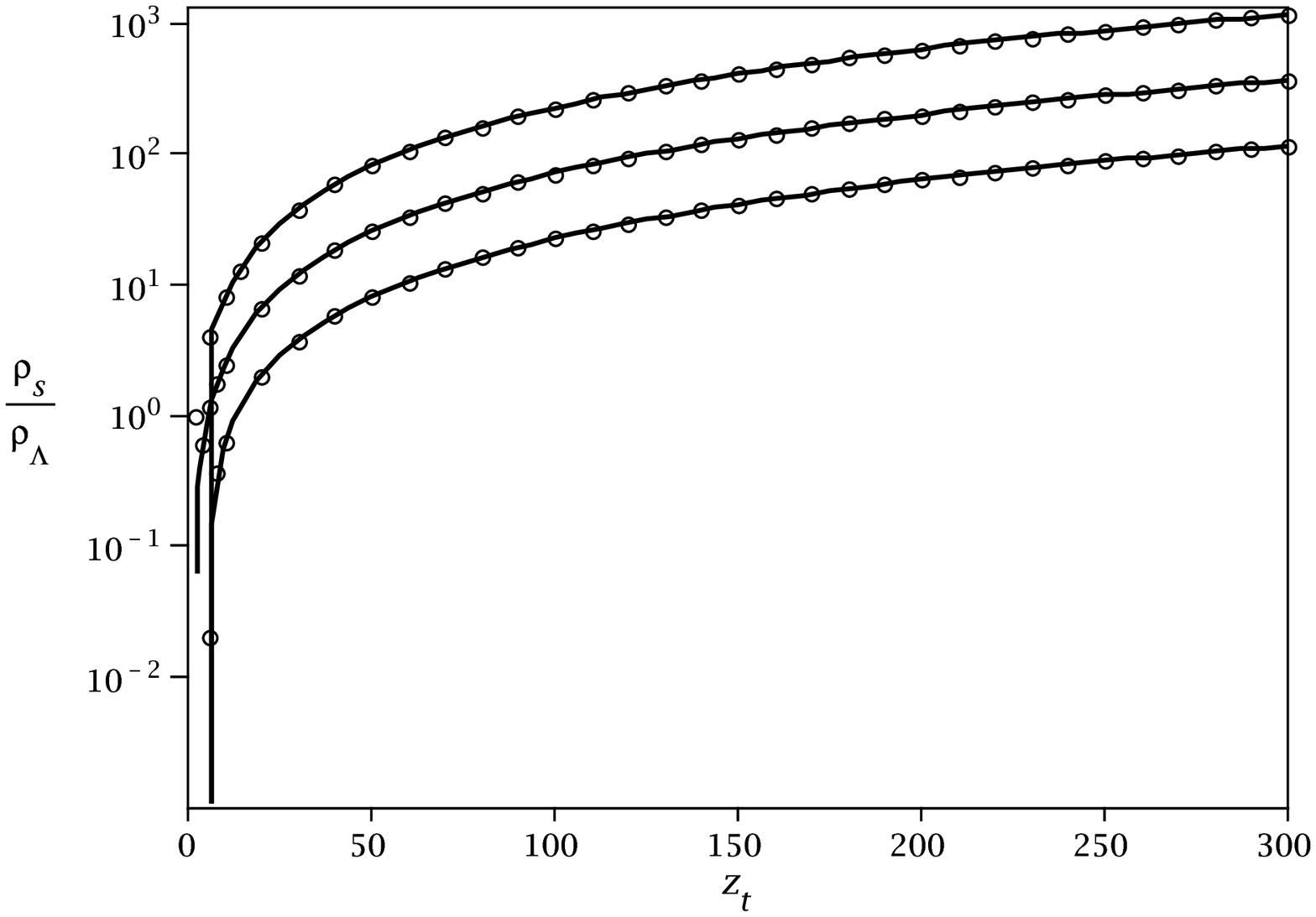}
\caption{The parameter $\rho_{\rm s}/\rho_\Lambda$ required to obtain a given $\hat{k}_{\rm J}$ (left panel) or a given  $E$ (right panel) {\it vs} the transition redshift $z_{\rm t}$. $\hat{k}_{\rm J}$ and $E$ are the values of the Jeans wave number and the efficiency at $\hat{\rho} \simeq \rho_{\rm t} - 0.66\rho_{\rm s}$, when ${d} c_{\rm s}^2/{d} \rho$ is maximum. Left panel: $\hat{k}_{\rm J} = 0.5, 1, 10$ $h$ Mpc$^{-1}$ from top to bottom. Right panel: $E = 10, 10^2, 10^3$ from top to bottom.  The solid lines represent the   theoretical approximations for  $\hat{k}_{\rm J}$ and $E$, the circles   the numerical values, see text.}
\label{rhosvszfkJmin}
\end{center}
\end{figure}

Evaluating the analytical expression (\ref{kJ2analytic}) of $k_{\rm J}(\rho)$ at $\hat{\rho}$  under the assumption $\rho_{\rm s} \ll \rho_{\rm t}$ we obtain the following approximate expression:
\begin{equation}\label{approxrelkjcapp}
 \hat{k}^2_{\rm J} \simeq \frac{\rho_{\rm t}}{4\left(1 + z_{\rm t}\right)^2}\frac{\rho_\Lambda}{\rho_{\rm s}}\left|6\left(\frac{\rho_{\rm s}}{\rho_\Lambda}\right)^2 + \frac{\rho_{\rm s}}{\rho_\Lambda} + 1 - 4\sqrt{3}\frac{\rho_{\rm t}}{\rho_\Lambda}\right|\;.
\end{equation}
Defining $\mathcal{D} := - \left[4\left(1 + z_{\rm t}\right)^2\hat{k}^2_{\rm J} +\rho_{\rm t}\right]/\rho_{\rm t}$ and making sure that we have $\rho\, {d} c_{\rm s}^2/{ d}\rho > \mathcal{B}$ for $\rho = \hat{\rho}$, we can then extract from (\ref{approxrelkjcapp}) the required relation between $\rho_{\rm s}$ and $\rho_{\rm t}$:
\begin{equation}\label{approxrelrhosrhof}
 \frac{\rho_{\rm s}}{\rho_\Lambda} = \frac{\mathcal{D} + \sqrt{96\sqrt{3}\frac{\rho_{\rm t}}{\rho_\Lambda} + \mathcal{D}^2 - 24}}{12}\;.
\end{equation}
In the left panel plots of Fig.\ \ref{rhosvszfkJmin}, we compare the analytical approximation (\ref{approxrelrhosrhof}) with the numerical calculations from Eq.\ (\ref{kJ2analytic}), for $\rho = \hat{\rho}$ and for $\hat{k}_{\rm J} = 0.5, 1, 10$ $h$ Mpc$^{-1}$, as functions of $z_{\rm t}$. The agreement between our analytical approximation and the numerical calculation is clearly very good. As we can see from the figure, if we require   larger values of $\hat{k}_{\rm J}$  then  $\rho_{\rm s}/\rho_\Lambda$ must be smaller, i.e.\ a faster transition is needed. On the other hand, if the transition takes place farther in the past, i.e.\ for increasing values of $z_{\rm t}$, this constraint is less stringent.

Having established a good approximation for $\hat{k}_{\rm J}$, we now want to determine for which values of $\rho_{\rm t}$ and $\rho_{\rm s}$ this quantity is well representative of  ${k}_{\rm J}$  around the transition, i.e.\ when we have a plateau as in Fig.\ \ref{comparison1}. In particular, this can be estimated from the difference of the values of $\rho\, { d} c_{\rm s}^2/{ d}\rho$ and $\mathcal{B}$ at $\rho = \hat{\rho}$. The larger is the difference, the faster is the transition and the higher is the plateau effect.
We therefore define the efficiency parameter $E := \rho \left({ d} c_{\rm s}^2/{ d}\rho\right) / \mathcal{B}|_{\rho = \hat{\rho}}$ which, under the assumption $\rho_{\rm t} \gg \rho_{\rm s}$, can be analytically approximated  from Eq.\ (\ref{kJ2analytic}):
\begin{equation}\label{approxE}
 E \simeq \frac{4\sqrt{3}\frac{\rho_{\rm t}}{\rho_\Lambda}}{6\left(\frac{\rho_{\rm s}}{\rho_\Lambda}\right)^2 + \frac{\rho_{\rm s}}{\rho_\Lambda} + 1}\;.
\end{equation}
From this, we obtain a new relation between $\rho_{\rm s}$ and $\rho_{\rm t}$:
\begin{equation}\label{approxrelrhosrhofeff}
 \frac{\rho_{\rm s}}{\rho_\Lambda} = \frac{1}{12}\frac{\sqrt{96\sqrt{3}E\frac{\rho_{\rm t}}{\rho_\Lambda} - 23E^2} - E}{E}\;.
\end{equation}
In the right panel of Fig.\ \ref{rhosvszfkJmin}, we compare the analytical approximation (\ref{approxrelrhosrhofeff}) with the numerical calculations, for $E = 10, 10^2, 10^3$, with very good  agreement. Notice that the larger is the efficiency $E$, the smaller  $\rho_{\rm s}/\rho_\Lambda$ must be, i.e.\ a faster transition is required.\footnote{Assuming  $\rho_{\rm s} \geq 0$  in Eq.\ (\ref{approxrelrhosrhofeff}) implies  $E\le 4\sqrt{3}~\left(\rho_{\rm t}/\rho_\Lambda\right)$. This limit in $E$ can be seen in the right panel of Fig.\ \ref{rhosvszfkJmin} and in Fig.\ \ref{cs2maxzfkJmin}, where in this limit $c_{\rm s\,Max}^{2} \to \infty$.}

It is important to stress that a large efficiency is relevant in order for $\hat{k}_{\rm J}$ to be a more  representative ``minimum on average'' value of ${k}_{\rm J}$ during the transition, i.e.\ it is not a necessary condition in order to have a fast transition or a model in good agreement with observation. This can be understood observing the multiplicative term $\left[3\rho(1 + w)\right]/\left[2(1 + z)^2c_{\rm s}^2\right]$ in front of the expression (\ref{kJ2analytic}) of the Jeans wave number  ${k}_{\rm J}$. Indeed, for increasing values of $\rho_{\rm t}$, this term amplifies the difference $\left(\rho d c_{\rm s}^2/d\rho\right) - \mathcal{B}$ during the transition, giving a larger ${k}_{\rm J}$. So, one can obtain models in good agreement with observation even if the efficiency is low.

Considering the plots in Figs.\ \ref{comparison3}-\ref{comparison1}, where for $\rho_{\rm s} = 10\rho_\Lambda, 10^{-1}\rho_\Lambda, 10^{-4}\rho_\Lambda$  we obtain respectively $E= 1.13, 597.26, 692.75$, in order to have a fast transition and a pronounced plateau, we infer that $E \gtrsim 700$ is needed. In addition, this requirement was also confirmed by the study of the matter power spectrum, see the next section. 

\begin{figure}[htbp]
\begin{center}
\includegraphics[width=0.6\columnwidth]{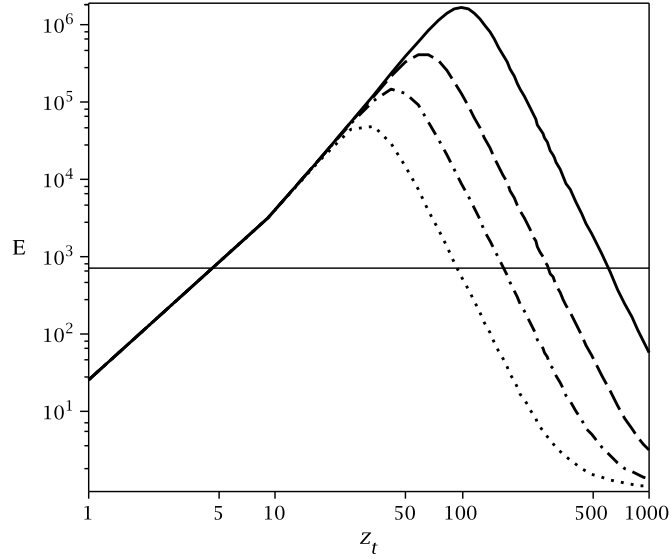}
\caption{Efficiency $E$ as function of $z_{\rm t}$, for $\hat{k}_{\rm J} = 5, 2, 1, 0.5$ $h$ Mpc$^{-1}$ (from top to bottom).}
\label{effzfkj}
\end{center}
\end{figure}

\begin{figure}[htbp]
\begin{center}
\includegraphics[width=0.8\columnwidth]{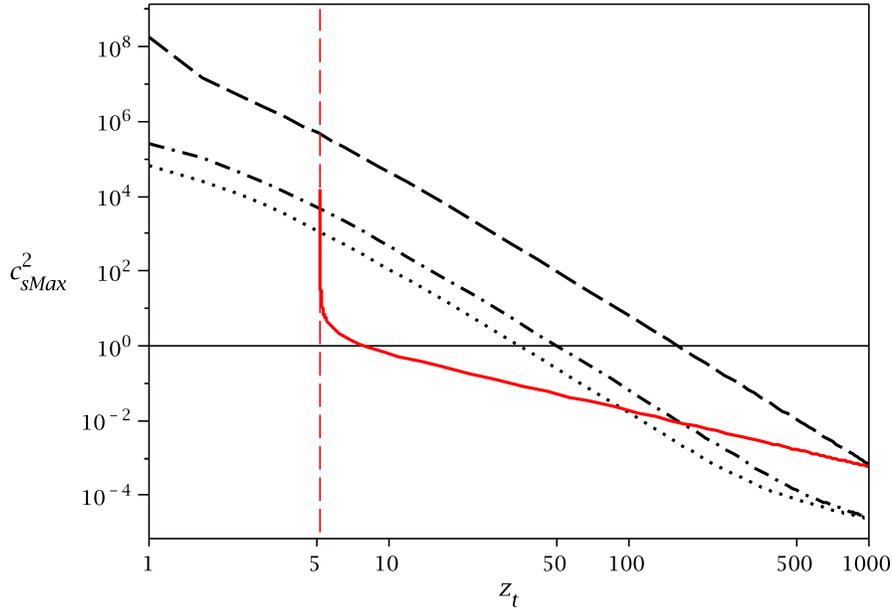}
\caption{
Evolution of the maximum value of $c_{\rm s}^2$ as function of the redshift $z_{\rm t}$. The choice of $\rho_{\rm s}$ as function of $\rho_{\rm t}$ is given by Eq.\ (\protect\ref{approxrelrhosrhof}) in the text for $\hat{k}_{\rm J} = 10, 1, 0.5$ $h$ Mpc$^{-1}$ (from top to bottom, dashed, dash-dotted and dotted line) and by Eq.\ (\protect\ref{approxE}) for $E = 700$ (solid red line).  
Then, 
the  vertical dashed line corresponds to the value of $z_{\rm t}$ for which $\rho_{\rm s} \to 0$ in Eq.\ (\protect\ref{approxrelrhosrhofeff}), giving $c_{\rm s\,Max}^{2} \to \infty$ in Eq.\ (\protect\ref{csmax}). }
\label{cs2maxzfkJmin}
\end{center}
\end{figure}
Substituting Eq.\ (\ref{approxrelrhosrhof}) in Eq.\ (\ref{approxE}), we can now obtain an approximate expression of $E$ as a function of the redshift of the transition $z_{\rm t}$, which allows us to estimate the range of $z_{\rm t}$ (and thus $\rho_{\rm t}$) for which the efficiency $E$ is above a certain threshold,  for a given  $\hat{k}_{\rm J}$. In Fig.\ \ref{effzfkj} we show the plot $E$ {\it vs} $z_{\rm t}$ for $\hat{k}_{\rm J} = 5, 2, 1, 0.5$ $h$ Mpc$^{-1}$ (from top to bottom).
For increasing values of $\hat{k}_{\rm J}$, the range of $z_{\rm t}$ in which $E > 700$ becomes larger, as expected.

 We can now use the relation (\ref{approxrelrhosrhof}) to understand how large the speed of sound can be during the transition. To this purpose, we substitute Eq.\ (\ref{approxrelrhosrhof}) in the maximum value of $c_{\rm s}^{2}$, at $\rho = \rho_{\rm t}$. We plot this $c_{\rm s\,Max}^{2}$ in Fig.\ \ref{cs2maxzfkJmin} as a function of $z_{\rm t}$ for $\hat{k}_{\rm J} = 10, 1, 0.5$ $h$ Mpc$^{-1}$ (from top to bottom). On the same figure we also plot  the curve of the maximum of the speed of sound  fixing the  value of the efficiency at $E = 700$ (solid red line). Then, 
the  vertical dashed line corresponds to the value of $z_{\rm t}$ for which $\rho_{\rm s} \to 0$ in Eq.\ (\protect\ref{approxrelrhosrhofeff}), giving $c_{\rm s\,Max}^{2} \to \infty$ in Eq.\ (\protect\ref{csmax}). 
In order to have $E > 700$, we must consider  the area above the solid red line. We can see that, for increasing values of $z_{\rm t}$, the value of $c_{\rm s\,Max}^{2}$ required to have a fixed $\hat{k}_{\rm J}$ decreases. For example, for $\hat{k}_{\rm J} = 1$ $h$ Mpc$^{-1}$, in order to have  $c_{\rm s\,Max}^{2} < 1$, the transition has to take place at $z_{\rm t} \gtrsim 50$,  while if $z_{\rm t} \simeq 5$ we have $c_{\rm s\,Max}^{2} \sim 10^{3}$.

\section{The CMB and matter Power spectra: toy model predictions}\label{sec:spectra}

In order to compare the predictions of our toy UDM model with observational data, we have used a properly modified version of CAMB\footnote{http://camb.info/} \cite{Lewis:1999bs} for the computation of the CMB and the matter power spectra. In particular, we have modified the original definition of the density contrast for the case of adiabatic UDM models.
Indeed, we have to define the UDM density contrast as $\delta := \delta\rho/\rho_{\rm m}$ \cite{Pietrobon:2008js}, where here $\rho_{\rm m}=\rho-\rho_{\Lambda}$ is the ``aether'' part of the UDM fluid \cite {Ananda:2005xp,Linder:2008ya}. In this case, starting from the perturbation theory we outlined in section \ref{sec:bgperteq}, we can infer the link between the density contrast and the gravitational potential via the Poisson equation in the following way:
\begin{equation}
 \delta\left(k;z\right) = \frac{\delta\rho\left(k;z\right)}{3H_0^2\Omega_{\rm m0}\left(1 + z\right)^3} = -k^{2}\frac{\Phi\left(1 + z\right)^2}{\left(3/2\right)\Omega_{\rm m0}}\;,
\end{equation}
for scales smaller than the cosmological horizon and $z < z_{\rm rec}$, where $z_{\rm rec}$ is the recombination redshift ($z_{\rm rec} \approx 10^{3}$).

We compare the theoretical predictions of our toy model with the WMAP 5-year data  \cite{Komatsu:2008hk,Dunkley:2008ie,Hinshaw:2008kr} and the luminous red galaxies power spectrum measured by the SDSS collaboration \cite{Tegmark:2006az}. The CMB data used in our plots are available on the LAMBDA\footnote{http://lambda.gsfc.nasa.gov} website, while those regarding the matter power spectrum are implemented in a modified version of the CosmoMC software\footnote{http://cosmologist.info/cosmomc/}.

We consider as reference the flat $\Lambda$CDM model described by the best-fit parameters found by combining  WMAP5 data  with measurements of Type Ia supernovae and Baryon Acoustic Oscillations \cite{Komatsu:2008hk,Hinshaw:2008kr}, with values provided on the LAMBDA website (68\% CL uncertainties): $\Omega_{\rm b0}h^2 = 0.02265\pm 0.00059$, $\Omega_{\rm m0}h^2 = 0.1143 \pm 0.0034$, $\Omega_\Lambda = 0.721 \pm 0.015$, $H_0 = 70.1  \pm 1.3 $ km s$^{-1}$ Mpc$^{-1}$, $n_{\rm s} = 0.960^{  + 0.014}_{ - 0.013}$ and $\Delta_{\rm R}^2 = (2.457^{ + 0.092}_{ - 0.093})\times10^{-9}$. For our toy model, we keep the same amount of baryons but choose a vanishing CDM content.

\begin{figure}[htbp]
\begin{center}
\includegraphics[width=0.75\columnwidth]{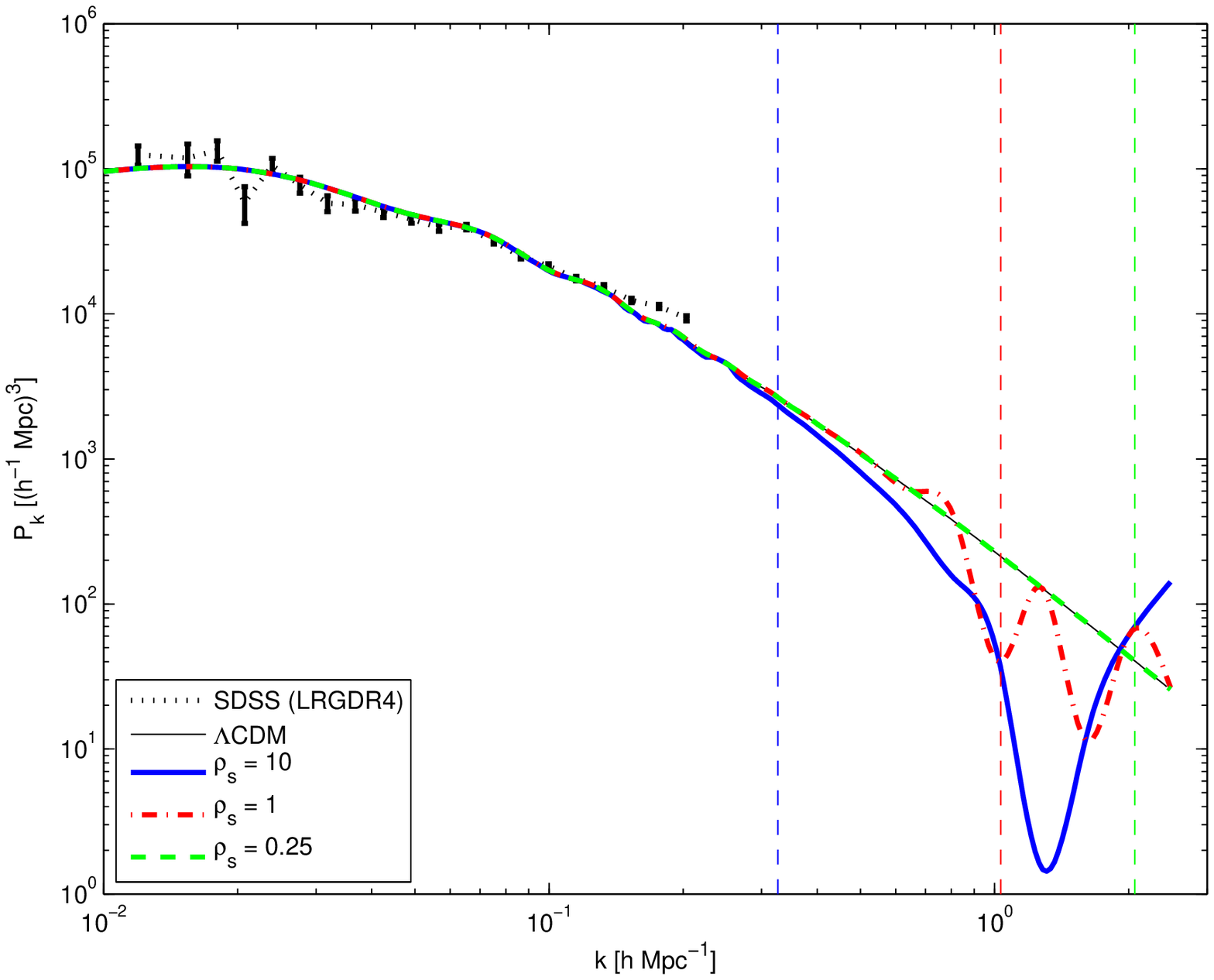}\\
\includegraphics[width=0.75\columnwidth]{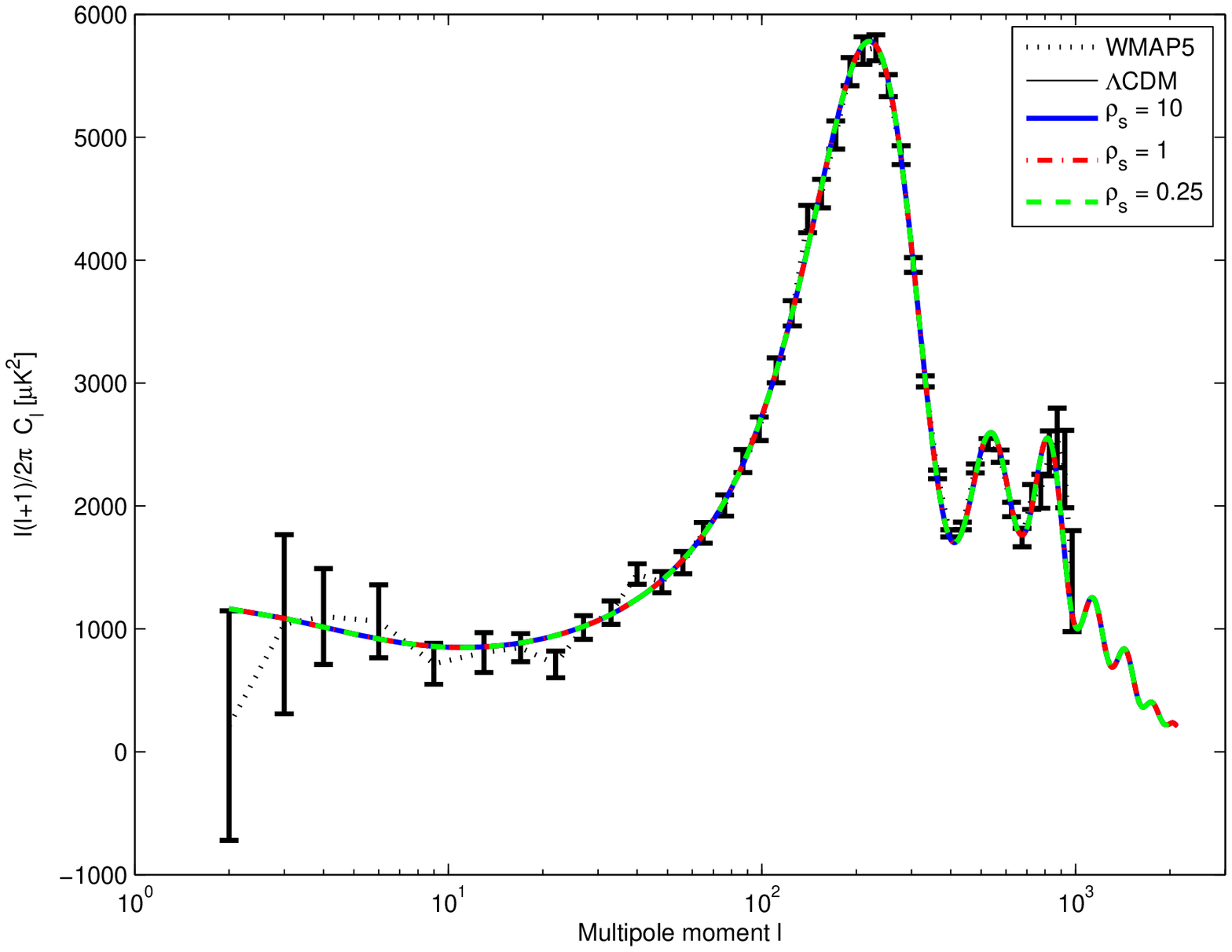}
\caption{Matter power spectrum (upper panel) and CMB power spectrum (lower panel) for   $\rho_{\rm t} = 10^{5}\rho_\Lambda$, i.e.\ a transition at $z_{\rm t} \sim 66$. 
The values of $\rho_{\rm s}/\rho_\Lambda$ are:
 $\rho_{\rm s}/\rho_\Lambda \simeq 0.25$ (dashed green line), $\rho_{\rm s}/\rho_\Lambda \simeq 1$ (dash-dotted red line) and $\rho_{\rm s}/\rho_\Lambda \simeq 10$ (solid blue line). The choice of the parameter has been done in order to have: $\hat{k}_{\rm J} \sim 2$ $h$ Mpc$^{-1}$ (dashed green vertical line) and $E \sim 4~10^5$ for $\rho_{\rm s}/\rho_\Lambda \simeq 0.25$; $\hat{k}_{\rm J} \sim 1$ $h$ Mpc$^{-1}$ (dashed red vertical line) and $E \sim 9~10^4$ for $\rho_{\rm s}/\rho_\Lambda \simeq 1$; $\hat{k}_{\rm J} \sim 0.3$ $h$ Mpc$^{-1}$ (dashed blue vertical line) and $E \sim 10^3$ for $\rho_{\rm s}/\rho_\Lambda \simeq 10$. The reference $\Lambda$CDM (see text) is represented by the solid black line. Notice that the theoretical curves representing the CMB power spectrum for our models and the reference $\Lambda$CDM are indistinguishable.}
\label{spectrarhof1e5}
\end{center}
\end{figure}

\begin{figure}[htbp]
\begin{center}
\includegraphics[width=0.73\columnwidth]{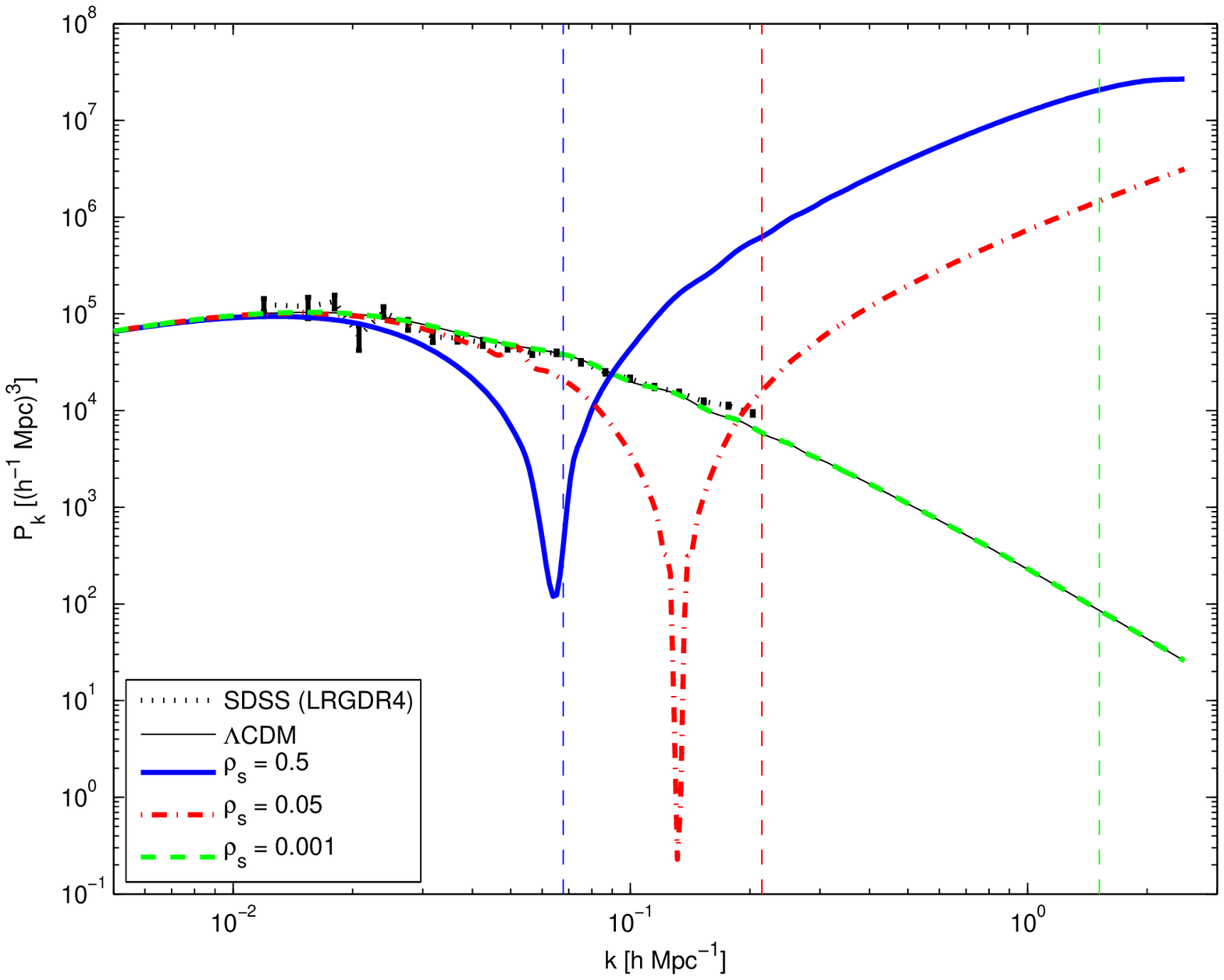}\\
\includegraphics[width=0.73\columnwidth]{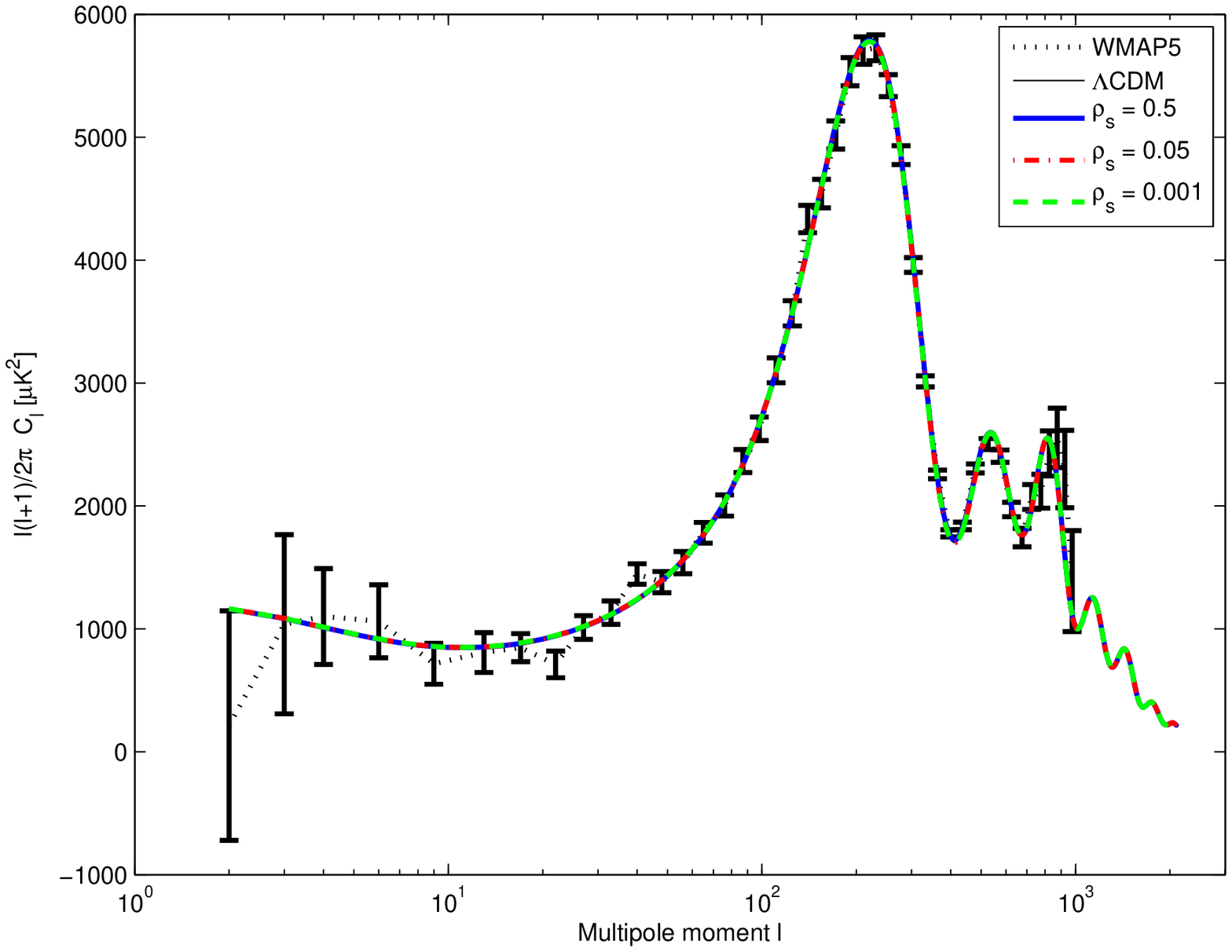}
\caption{Matter power spectrum (upper panel) and CMB power spectrum (lower panel) for  $\rho_{\rm t} = 10^{3}\rho_\Lambda$, i.e.\  $z_{\rm t} \sim 13$.
The values of $\rho_{\rm s}/\rho_\Lambda$ are:
$\rho_{\rm s}/\rho_\Lambda \simeq 10^{-3}$ (dashed green line), $\rho_{\rm s}/\rho_\Lambda \simeq 0.05$ (dash-dotted red line) and $\rho_{\rm s}/\rho_\Lambda \simeq 0.5$ (solid blue line). The choice of the parameter has been done in order to have: $\hat{k}_{\rm J} \sim 1.5$ $h$ Mpc$^{-1}$ (dashed green vertical line) and $E \sim 7~10^3$ for $\rho_{\rm s}/\rho_\Lambda \simeq 10^{-3}$; $\hat{k}_{\rm J} \sim 0.2$ $h$ Mpc$^{-1}$ (dashed red vertical line) and $E \sim 7~10^3$ for $\rho_{\rm s}/\rho_\Lambda \simeq 0.05$; $\hat{k}_{\rm J} \sim 0.07$ $h$ Mpc$^{-1}$ (dashed blue vertical line) and $E \sim 2~10^3$ for $\rho_{\rm s}/\rho_\Lambda \simeq 0.5$. Also in this case $z_{\rm t} \sim 13$ the theoretical curves representing the CMB power spectrum for our models and the reference $\Lambda$CDM are indistinguishable. However, the matter power spectrum requires a faster transition, i.e.\ smaller $\rho_{\rm s}$ values.}
\label{spectrarhof1e3}
\end{center}
\end{figure}

\begin{figure}[htbp]
\begin{center}
\includegraphics[width=0.75\columnwidth]{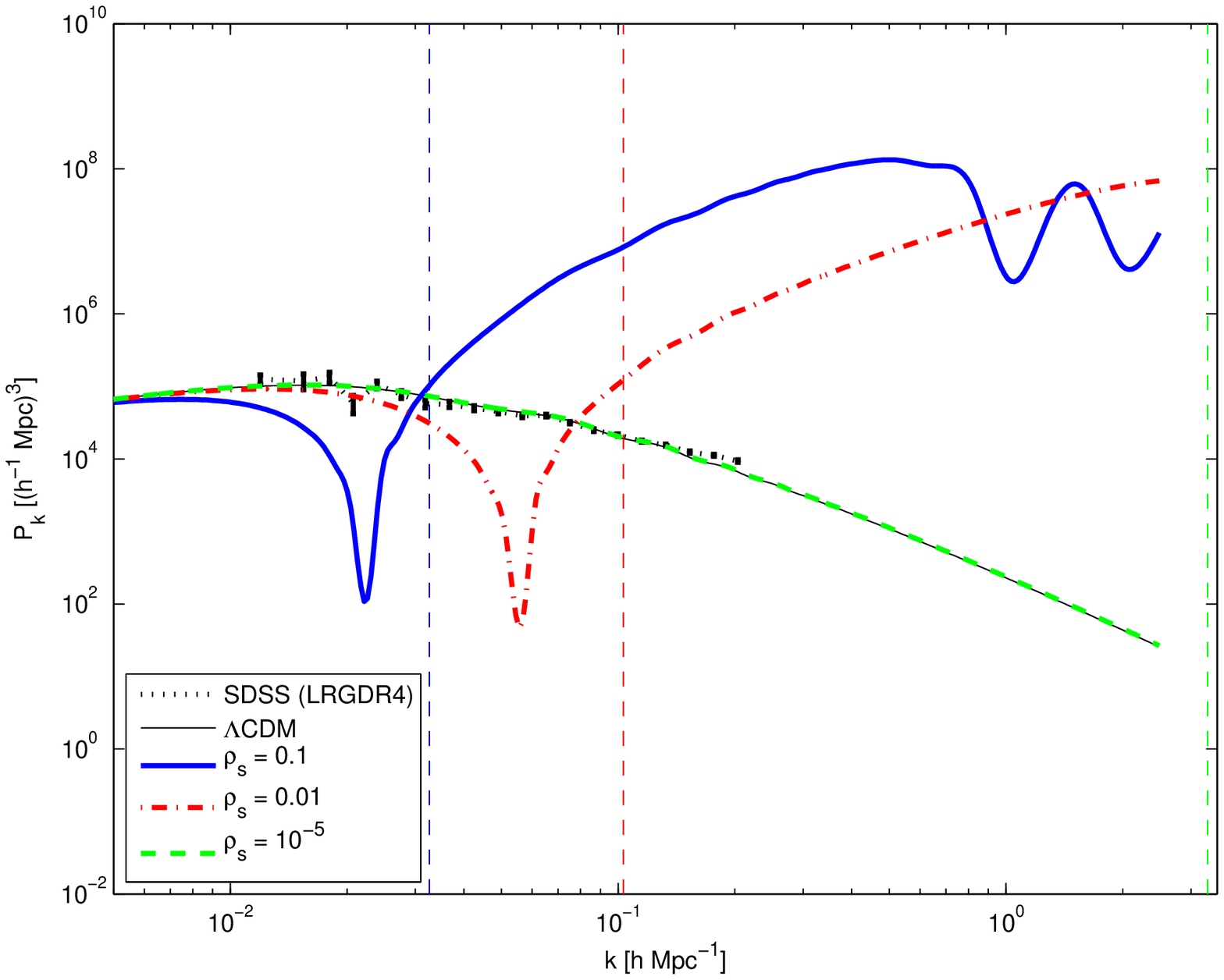}\\
\includegraphics[width=0.75\columnwidth]{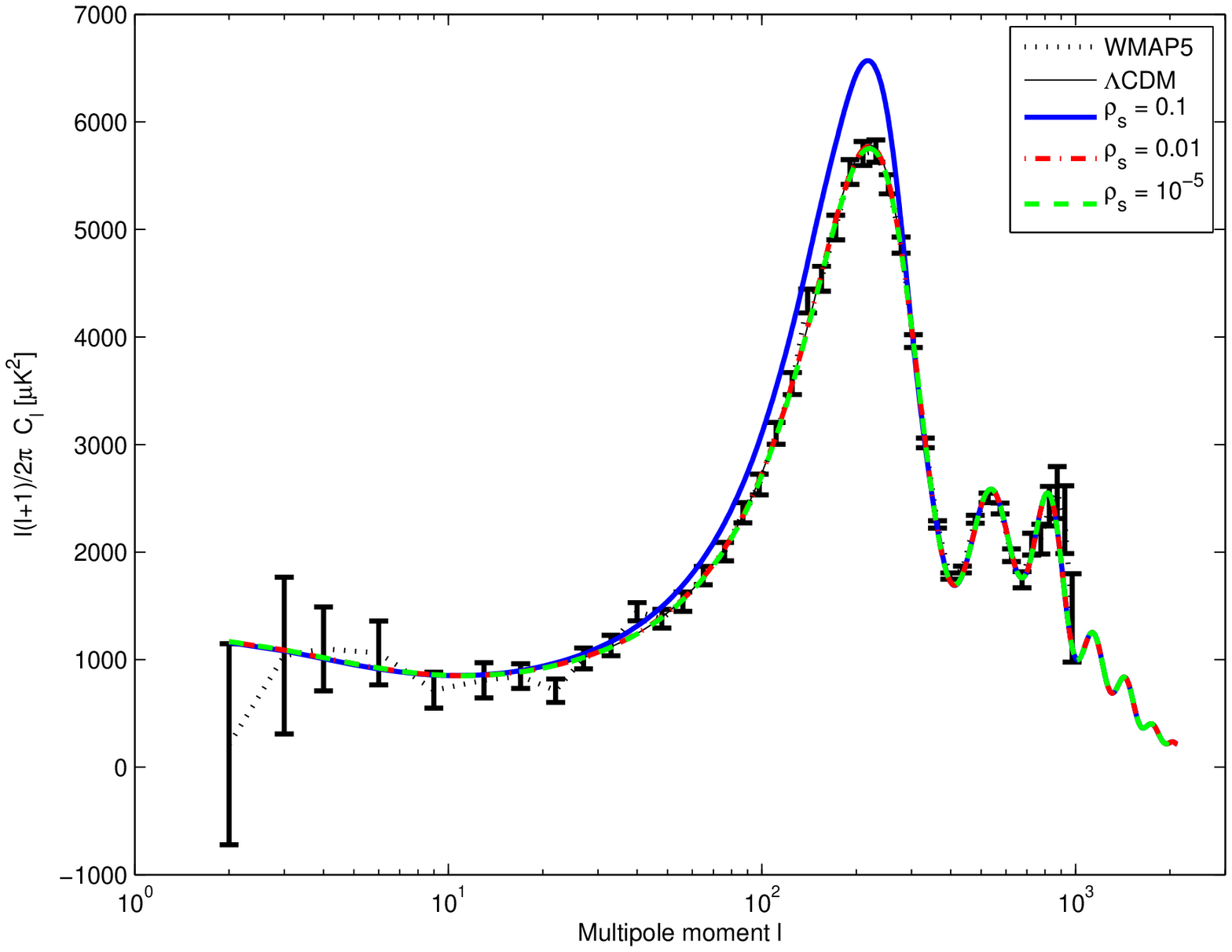}
\caption{Matter power spectrum (upper panel) and CMB power spectrum (lower panel) for  $\rho_{\rm t} = 10^{2}\rho_\Lambda$, i.e.\  $z_{\rm t} \sim 5.7$.
The values of $\rho_{\rm s}/\rho_\Lambda$ are:
 $\rho_{\rm s}/\rho_\Lambda \simeq 10^{-5}$ (dashed green line), $\rho_{\rm s}/\rho_\Lambda \simeq 0.01$ (dash-dotted red line) and $\rho_{\rm s}/\rho_\Lambda \simeq 0.1$ (solid blue line). The choice of the parameter has been done in order to have: $\hat{k}_{\rm J} \sim 3.3$ $h$ Mpc$^{-1}$ (dashed green vertical line) and $E \sim 6.9~10^2$ for $\rho_{\rm s}/\rho_\Lambda \simeq 10^{-5}$; $\hat{k}_{\rm J} \sim 0.1$ $h$ Mpc$^{-1}$ (dashed red vertical line) and $E \sim 6.9~10^2$ for $\rho_{\rm s}/\rho_\Lambda \simeq 0.01$; $\hat{k}_{\rm J} \sim 0.03$ $h$ Mpc$^{-1}$ (dashed blue vertical line) and $E \sim 6~10^2$ for $\rho_{\rm s}/\rho_\Lambda \simeq 0.1$. At the relatively small transition redshift  $z_{\rm t} \sim 5.7$ a viable matter power spectrum requires an even faster transition, i.e.\  smaller $\rho_{\rm s}$ values. }
\label{spectrarhof1e2}
\end{center}
\end{figure}

\begin{figure}[htbp]
\begin{center}
\includegraphics[width=0.71\columnwidth]{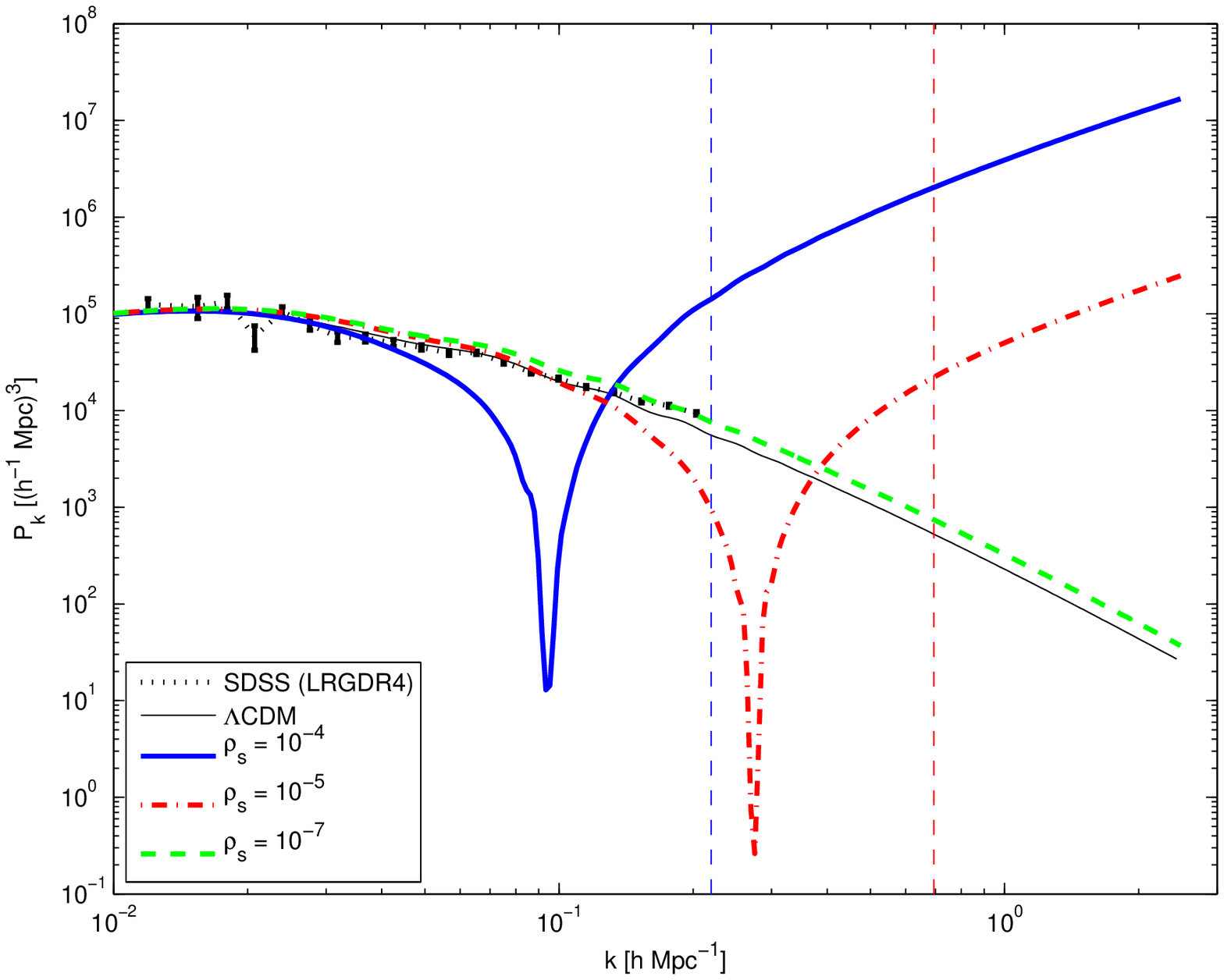}\\
\includegraphics[width=0.71\columnwidth]{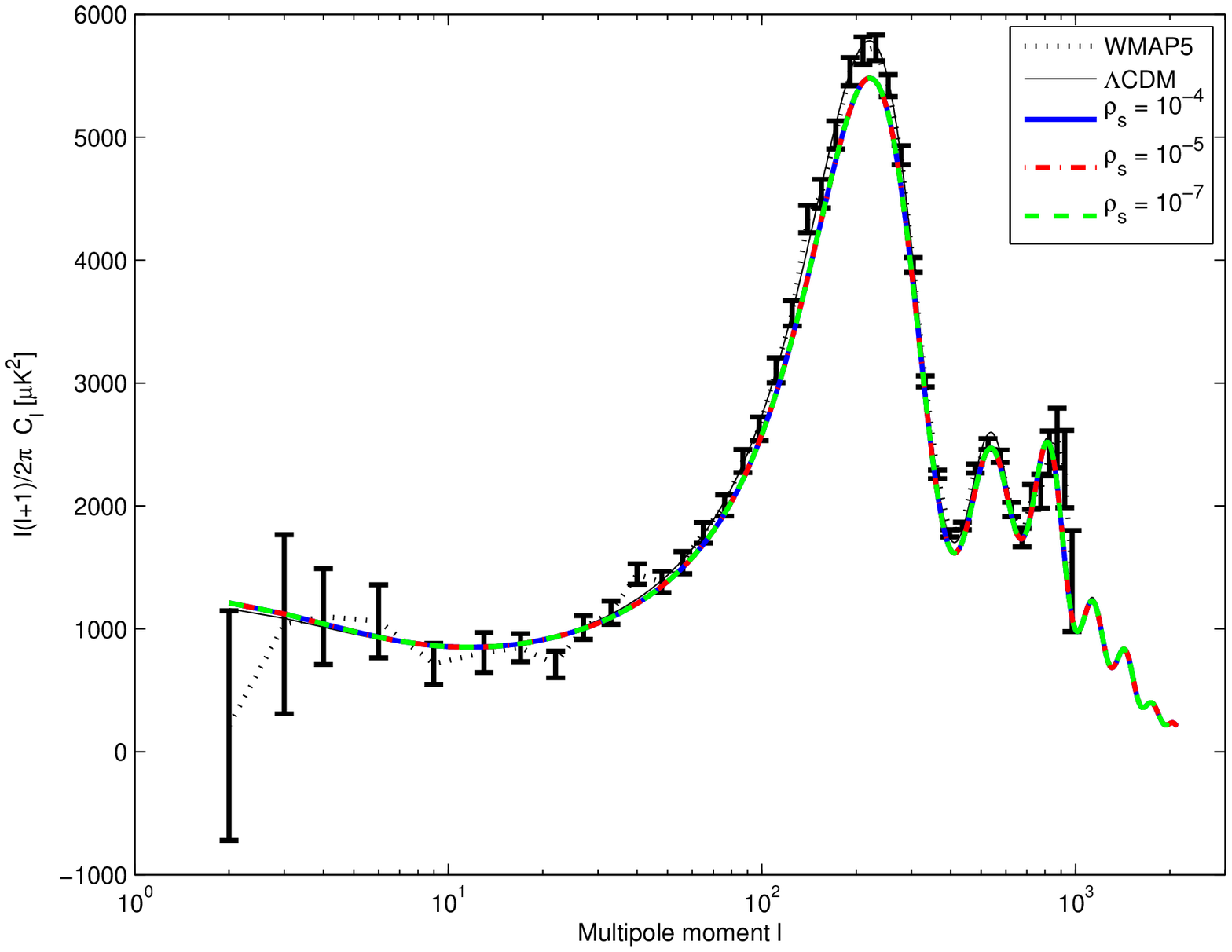}
\caption{Matter power spectrum (upper panel) and CMB power spectrum (lower panel) for $\rho_{\rm t} = 10\rho_\Lambda$, i.e.\ $z_{\rm t} \sim 2$. 
The values of $\rho_{\rm s}/\rho_\Lambda$ are:
 $\rho_{\rm s}/\rho_\Lambda \simeq 10^{-7}$ (dashed green line), $\rho_{\rm s}/\rho_\Lambda \simeq 10^{-5}$ (dash-dotted red line) and $\rho_{\rm s}/\rho_\Lambda \simeq 10^{-4}$ (solid blue line). The choice of the parameter has been done in order to have: $\hat{k}_{\rm J} \sim 7$ $h$ Mpc$^{-1}$ (dashed green vertical line) and $E \sim 70$ for $\rho_{\rm s}/\rho_\Lambda \simeq 10^{-7}$; $\hat{k}_{\rm J} \sim 0.7$ $h$ Mpc$^{-1}$ (dashed red vertical line) and $E \sim 68$ for $\rho_{\rm s}/\rho_\Lambda \simeq 10^{-5}$; $\hat{k}_{\rm J} \sim 0.23$ $h$ Mpc$^{-1}$ (dashed blue vertical line) and $E \sim 68$ for $\rho_{\rm s}/\rho_\Lambda \simeq 10^{-4}$. At this smaller transition redshift $z_{\rm t} \sim 2$ the background evolution in our model is so strongly modified that, no matter how fast the transition is, i.e.\ even for $\rho_{\rm s}$ values giving an acceptable matter power spectrum, there is no way to fit the CMB first peak for the given choice of $\Omega_\Lambda$, i.e.\ the same than the reference $\Lambda$CDM (see text).}
\label{spectrarhof1e1}
\end{center}
\end{figure}

In Figs.\ \ref{spectrarhof1e5}-\ref{spectrarhof1e1} we plot the theoretical predictions of our model, those of the reference $\Lambda$CDM and the observed CMB and matter power spectra data. Each of Figs.\ \ref{spectrarhof1e5}-\ref{spectrarhof1e1} respectively correspond to  $\rho_{\rm t}/\rho_\Lambda = 10^{5}, 10^{3}, 10^{2}, 10$, i.e.\ to $z_{\rm t} \simeq 66, 13, 5.7, 2$. Guided by the analysis in section \ref{sec:perts}, for each transition density $\rho_{\rm t}$ we have chosen values of $\rho_{\rm s}$ which clearly show the progressive enhancement of the agreement between the predicted matter power spectrum and the observed one. Moreover, in the matter power spectrum plots, for each choice $(\rho_{\rm t},\rho_{\rm s})$ we draw a vertical dashed line representing the corresponding value of $\hat{k}_{\rm J}$.

We can see from Figs.\ \ref{spectrarhof1e5} and \ref{spectrarhof1e3} that the CMB anisotropies predicted by the reference $\Lambda$CDM and by our  toy model are indistinguishable for a large range of $\rho_s$. However, while at the higher transition redshift of Fig.\ \ref{spectrarhof1e5}  the matter power spectrum also allows the same broad range of $\rho_{\rm s}$ values, at the smaller $z_{\rm t} $ of Fig.\  \ref{spectrarhof1e3} we start to see the need for a faster transition, i.e.\ a smaller  $\rho_{\rm s}$, to have an acceptable power spectrum. As expected from the analysis of  section \ref{sec:perts}, the effect becomes  more pronounced as the transition occurs at the  smaller and smaller redshifts of  Figs.\ \ref{spectrarhof1e2} and  \ref{spectrarhof1e1}.
In the case $\rho_{\rm s}/\rho_\Lambda \simeq 0.1$ of Fig.\ \ref{spectrarhof1e2} the first acoustic peak of CMB is higher with respect to the observational data. This effect can be explained by looking at the matter power spectrum. Indeed, the latter moves away from the reference $\Lambda$CDM  before the equivalence wavenumber $k_{\rm eq} \approx 0.01$ $h$ Mpc$^{-1}$. In other words, the gravitational potential starts to oscillate and to decay for $k < k_{\rm eq}$, therefore affecting those modes entering the horizon before the matter-radiation equivalence epoch.

\begin{figure}[htbp]
\begin{center}
\includegraphics[width=0.75\columnwidth]{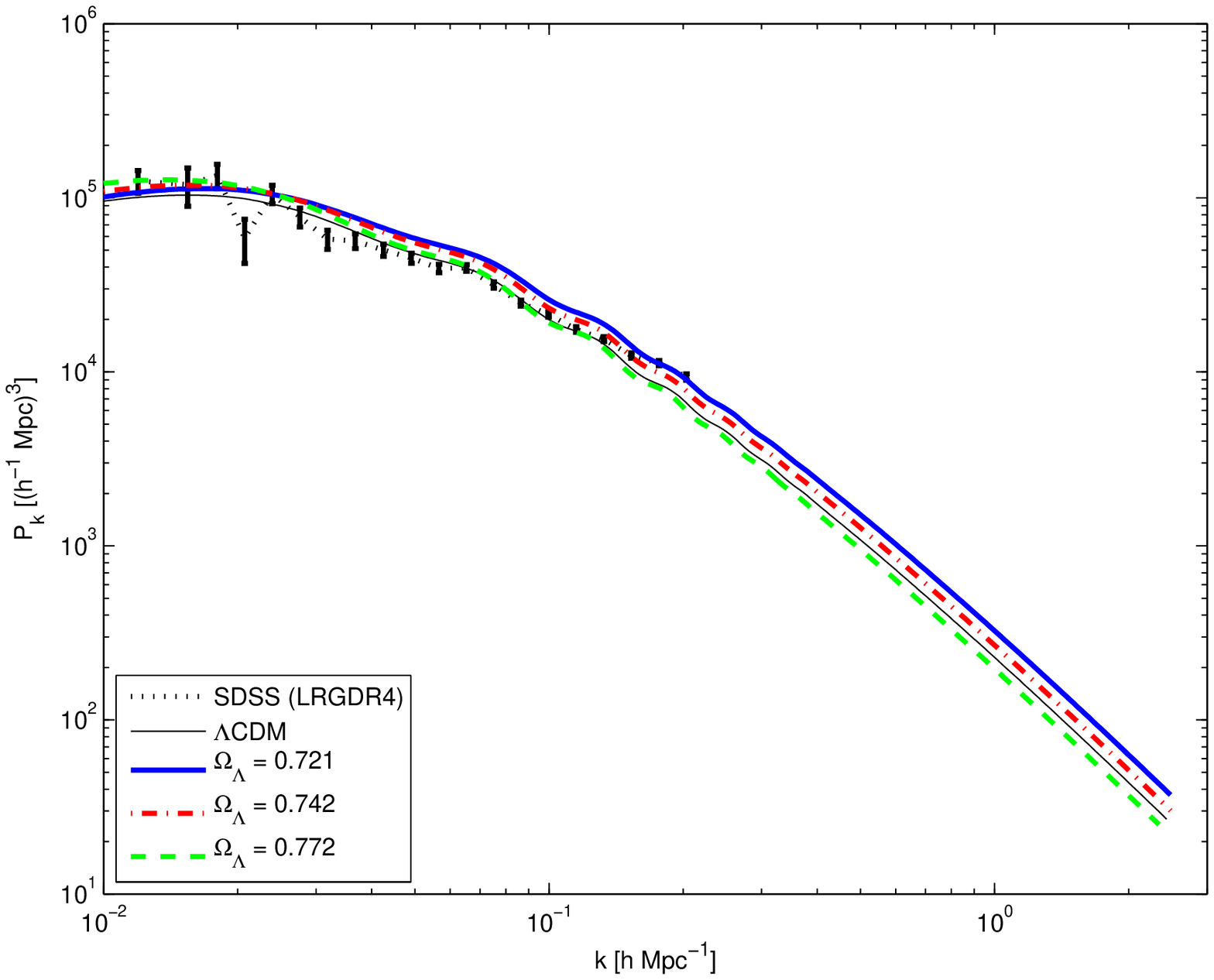}\\
\includegraphics[width=0.75\columnwidth]{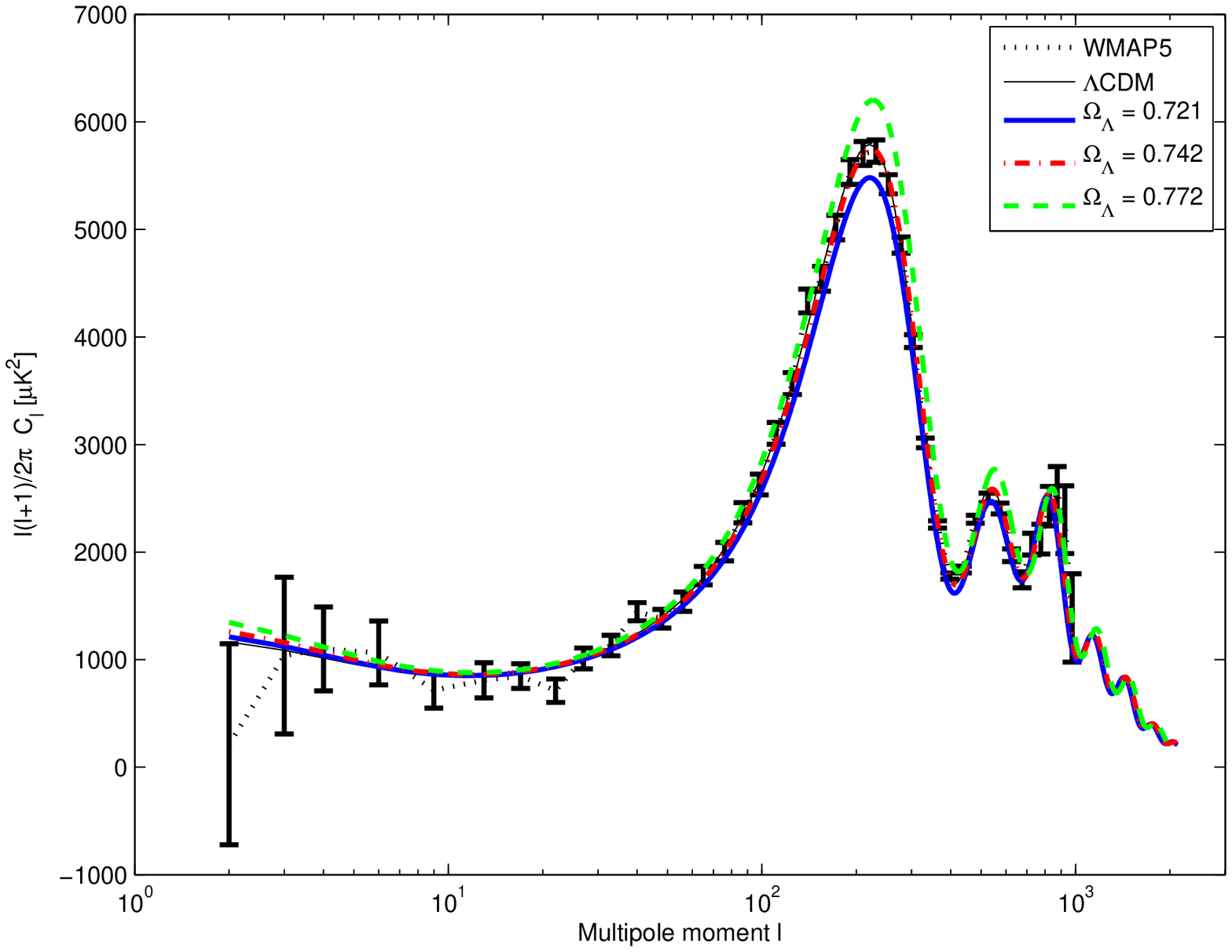}
\caption{Matter power spectrum (upper panel) and CMB power spectrum (lower panel) for one of the transition model of Fig.\ \protect\ref{spectrarhof1e1}, i.e.\  $\rho_{\rm t} = 10\rho_\Lambda$ and $\rho_{\rm s}/\rho_\Lambda \simeq 10^{-7}$, with a transition at   $z_{\rm t} \sim 2$, this time for different values of $\Omega_\Lambda$. The solid blue, red dot-dashed and green dashed lines respectively correspond to $\Omega_\Lambda =0.721, 0.742, 0.772$. The reference  $\Lambda$CDM (solid black line) is the same of the other figures. With a slightly higher $\Omega_\Lambda=0.742$ (WMAP5 best fit value with CMB data alone \cite{Dunkley:2008ie}) our  model now  produces an acceptable fit.  }
\label{rhot10rhos1em5Lambdavar}
\end{center}
\end{figure}

Finally, in Fig.\ \ref{spectrarhof1e1} the first CMB spectrum peak is lower than the observed one for any value of $\rho_{\rm s}$. Note in the left panel of Fig.\ \ref{wtanhzFig2} that, for $\rho_{\rm t} = 10\rho_\Lambda$, the background evolution is sensibly different from the reference $\Lambda$CDM. Indeed, in this case  our model behaves like a pure CDM Einstein--de\! Sitter for a much longer time. One possibility to avoid this discrepancy is to slightly increase $\Omega_\Lambda$.
Therefore, again using $\rho_{\rm t} = 10\rho_\Lambda$ and $\rho_{\rm s} = 10^{-5}\rho_\Lambda$, in Fig.\ \ref{rhot10rhos1em5Lambdavar} we have chosen $\Omega_\Lambda = 0.721, 0.742, 0.772$, with the first value again corresponding to the above mentioned reference $\Lambda$CDM and the other two corresponding to the best-fit and its upper uncertainty obtained by WMAP5 using CMB data only (see \cite{Dunkley:2008ie} and the LAMBDA website). The agreement between the CMB prediction and the observational data is again good for $\Omega_\Lambda = 0.742$, with a good matter power spectrum. 


\section{Conclusions}\label{sec:concl}


The last decade of observations of large scale structure \cite{Allen:2004cd,Tegmark:2006az,Percival:2006kh,Percival:2007yw,Percival:2009xn,Reid:2009xm}, the search for
Ia supernovae (SNIa) \cite{Riess:1998cb,Perlmutter:1998np,Riess:1998dv,Kowalski:2008ez} and the measurements of the CMB anisotropies
\cite{Spergel:2003cb,Komatsu:2008hk,Dunkley:2008ie,Hinshaw:2008kr} are very well explained by assuming  that two dark
components govern the dynamics of the Universe. They are DM, thought to be
the main responsible for structure formation, and an additional DE component
that is supposed to drive the measured cosmic acceleration \cite{Copeland:2006wr,Peebles:2002gy,Padmanabhan:2002ji}.

At the same time, in the context of General Relativity,
it is very interesting to study  possible alternatives. A popular one is that of an interaction between DM
and DE,  without violating current observational constraints \cite{Copeland:2006wr,Quercellini:2008vh,Valiviita:2008iv,CalderaCabral:2008bx,Majerotto:2009np,CalderaCabral:2009ja,Valiviita:2009nu}.
This possibility  could alleviate the so called ``coincidence problem" \cite{Zlatev:1998tr}, namely, why are the
energy densities of the two dark components of the same order of magnitude today.
Another attractive, albeit radical, explanation of  the observed cosmic
acceleration and structure formation  is to assume the existence
of a single dark component: UDM models 
\cite{Kamenshchik:2001cp,Bilic:2001cg, Sandvik:2002jz,Carturan:2002si, Amendola:2003bz,Makler:2003iw, Scherrer:2004au, Giannakis:2005kr, Bertacca:2007cv, Bertacca:2007ux, Bertacca:2007fc, Quercellini:2007ht, Bertacca:2008uf, Balbi:2007mz,Pietrobon:2008js, Camera:2009uz} where, by definition, there is no coincidence problem.

In the present paper we have investigated the general properties of UDM fluid models where the pressure and the energy density are linked by a barotropic equation of state (EoS) $p=p(\rho)$ and the perturbations are adiabatic.
Using the pressure-density plane, we have analysed  the properties that a general barotropic UDM model has to fulfil in order to be viable. We have assumed that the EoS of UDM models admits a future attractor which acts as an effective cosmological constant, while asymptotically
in the past the pressure is negligible, studying the possibility of constructing adiabatic UDM models where the Jeans length is very small, even when the speed of sound $c_{\rm s}$ is not negligible. In particular, we have focused on models that admit an effective cosmological
constant and that are characterised by a short period during which the effective speed of sound varies significantly from zero. 
This allows a fast transition between an early epoch that is indistinguishable from a standard matter dominated era, i.e.\ an Einstein de~Sitter model, and a more recent epoch whose dynamics,
background and perturbative, are very close to that of a standard $\Lambda$CDM model.

In the second part  of the paper, in order to quantitatively investigate observational
constraints on UDM models with fast transition, we have introduced and discussed a  toy model based on a hyperbolic tangent EoS [see  Eq.\ (\ref{tanh})]. We have shown that if the  transition takes place early enough,  at a redshift $z_{\rm t}\ \gtrsim 2$ when the effective cosmological constant is still subdominant,  being also fast enough, then these models can avoid the oscillating and decaying time evolution of the gravitational potential  that in many UDM models causes  CMB and matter fluctuations incompatible with observations. Consequently, the background evolution, the CMB anisotropy and the linear matter power spectrum predicted by our model do not display significant differences from those computed in a reference   $\Lambda$CDM  \cite{Komatsu:2008hk,Hinshaw:2008kr}, because the Jeans length $\lambda_{\rm J}=a/k_{\rm J}$, where $k_{\rm J}$ is the Jeans wave number [see Eq.\ (\ref{kJ2analytic})], remains  small at all times, except for negligibly short periods, even if during the fast transition the speed of sound can be large. In this way, our toy models (and more in general UDM models with a similar fast transition) can evade the ``no-go theorem'' of Sandvik {\it et al} \cite{Sandvik:2002jz}, as we discussed in the introduction.

Specifically, we have analysed the properties of perturbations in our toy model, focusing on the the evolution of the effective speed of sound  and that of the Jeans length during the transition. In this way, we have been able to set theoretical constraints on the parameters of the model, predicting sufficient conditions for the model
to be viable.
Finally, guided by  these predictions  and using  the CAMB code \cite{Lewis:1999bs}, we have computed the CMB and
the matter power spectra  showing that our toy model, for a wide range  of parameters values, fits observation. 

The full likelihood analysis for this model and its parameters would be an interesting extension of the study carried out here, which we will address in a future work.

\acknowledgments{OFP and DB wish to thank the ICG Portsmouth for the hospitality during the development of this project. The authors also thank  N.\ Bartolo, B.\ Bassett,  R.\ Crittenden, R.\ Maartens, S.\ Matarrese, S.\ Mollerach, M.\ Sasaki and M.\ Viel for discussions and suggestions.  DB research has been  partly supported by ASI  contract I/016/07/0 ``COFIS".}

\bibliographystyle{JHEP}
\bibliography{Bexp}

\end{document}